\documentclass[draft,showpacs,preprintnumbers,amsmath,amssymb]{revtex4}
\newcommand{\fant}[1]{\phantom{#1}}
\newcommand{\be}{\begin{equation}}
\newcommand{\ee}{\end{equation}}
\newcommand{\wdg}{\wedge}
\newcommand{\ot}{\otimes}

\begin{document}

\begin{abstract}
Our main aim in this paper is to promote the coframe variational method  as  a unified approach to derive field equations for any given gravitational action containing the algebraic functions of the scalars constructed from the Riemann curvature tensor and its contractions. We are able to derive a master equation  which expresses the variational derivatives of the generalized gravitational actions in terms of the variational derivatives of its constituent  curvature scalars. Using the Lagrange multiplier method relative to an orthonormal coframe, we investigate the variational procedures for modified gravitational
Lagrangian densities  in  spacetime dimensions $n\geqslant 3$. We study well-known gravitational actions such as those involving the Gauss-Bonnet  and Ricci-squared, Kretchmann scalar, Weyl-squared  terms and their algebraic generalizations similar to  generic $f(R)$ theories and the algebraic generalization of sixth order gravitational Lagrangians. We put forth a new model involving the gravitational Chern-Simons term and also give three dimensional  New massive gravity equations in a new form in terms of the Cotton 2-form.
\end{abstract}

\title{A Unified Approach to Variational Derivatives of  Modified  Gravitational Actions}
\pacs{04.20.Fy, 04.50.-h, 04.50.Kd, 11.15.Yc, 11.15.Wx}

\date{\today}

\author{ Ahmet Baykal}
\email{abaykal@nigde.edu.tr}
\affiliation{Department of Physics, Faculty of Science and Letters, Ni\u gde University,  51240, Ni\u gde, Turkey}

\author{\"Ozg\"ur Delice}
\email{ozgur.delice@marmara.edu.tr}
\affiliation{Department of Physics, Faculty of Science and Letters, Marmara University, 34722, Istanbul, Turkey}

\maketitle

\section{Introduction}

Although General Relativity theory is an  experimentally well-tested and successful theory \cite{MTW}, the search for alternative or modified gravity theories emerged immediately after the formulation of General Relativity (GR). The motivation of introducing such theories varies considerably. For example, in some of the  earlier theories, such as those introduced by Kretschmann \cite{Kretschmann}, Weyl \cite{Weyl} and Eddington \cite{Eddington},  Lagrangians containing higher order curvature scalars  were considered in order to understand mathematical and physical machinery of Einstein's theory and to compare their predictions with different and more generalized theories. In another important route, in the same era, Kaluza and Klein showed that it is possible to unify  gravity with electromagnetism from pure geometric Einstein-Hilbert (EH) action but employing an extra compact dimension \cite{Kaluza}. Although this theory
is not a viable theory, its main idea of employing extra dimensions for the unification survive in modern theories such as String/M theory \cite{String}. In these theories, terms involving higher order curvature scalars arise naturally in gravitational sector of their action \cite{Stringcorrections}. It is also known that, in order to incorporate quantum effects to gravity, one might employ Wheeler-DeWitt effective action formalism \cite{deWitt}, which leads to higher order curvature corrections as well \cite{deWittcorrections}. Hence, the studies considering unification as well as quantization of gravity require modifications of some sort in the gravitational Lagrangian.

Another motivation for considering modified gravitational actions comes from the dynamics of the universe. The results of data accumulated on cosmology and astrophysics, in recent years, lead to interesting observations, for example, that the universe is in the epoch of accelerating expansion driven by
the dark energy  and most of the matter content of the universe is dark \cite{darkenergy}. Although the ongoing and future particle physics experiments may explain the physical content of dark matter and dark energy, the possibility that these observations suggests that GR should be modified
is also an interesting and attractive idea. Hence,  in recent years, the old idea of replacing $R$ in EH action with  $f(R)$ \cite{bergman,ruzmaikina,breizman,buchdahl-first-fr,starobinsky}, has been studied extensively in this context. For more information as well as recent developments and their applications to the physically relevant models of $f(R)$ theories, see one of the excellent reviews \cite{Faraoni,felice,querella} and references therein. The problem of this approach is that, can these modified gravity theories explain those observational phenomena without bringing
out more problems to the area, such as instabilities, and can be compatible with laboratory and observational astrophysical data.  Actually, one should regard these attempts  as toy models, which might help to understand the gravitational aspects of nature by modeling and studying geometric gravity theories.
It is preferable to start investigating modified  gravitational models by starting from   action scalar for which the metric field equations derived from an action via a well-defined variational procedure. There are  distinct variational procedures that lead to distinct field equations. In so-called Palatini formalism, the metric tensor and connection are treated as independent variables of the theory and the relation between them is derived as a consequence. In contrast, in the metric formalism, the metric tensor is assumed to be the only field variable determining the spacetime geometry and the field equations for it follows from the variational derivatives of the gravitational action with respect to the components of the metric tensor. Except for the Einstein-Hilbert
Lagrangian the metric field equations that follow from these two procedure, as we shall explicitly show below, are not the same. In addition, in the metric approach, the connection (Riemannian or Levi-Civita connection) is derived uniquely from the derivatives of the metric tensor, whereas in the Palatini approach, the connection may be non-Riemannian. This has important consequences on the motions of test particles since a connection defines parallel transport and therefore determine the geodesic structure of the spacetime.

The organization of our paper can be outlined as follows. The paper can roughly be divided into three parts. The first part deals with the constrained variational procedure and Lagrange multiplier technique that will be used in the rest of the paper. The second part is  related to the algebraic generalization of modified gravitational actions. In this part, the variational derivative  of the well-known $f(R)$ theory in both metric and Palatini approaches
and the equivalence of these actions to the Brans-Dicke (BD) type scalar tensor theories are discussed. In the subsequent sections we calculate the variational derivative of the actions which consist of functions of the terms such as Gauss-Bonnet, Ricci-squared or Weyl-squared terms and their algebraic generalizations. Some of the results related to the algebraic generalization of the quadratic curvature terms are new. Various important inter-relationships between the Lagrangians in three and four dimensions are indicated. Finally, the variation of terms involving derivatives of the curvature scalars, which lead to
sixth order theories, as well as their algebraic generalizations are  discussed. The third part is the original part and  it consist of last two sections.
These sections can be considered as applications of the general formulae obtained in the previous parts. In the first of these sections, namely section V, a new model  for the Chern-Simons gravity in four dimensions is derived whereas  the last section is devoted to the study of  field equations of three dimensional general massive  gravity models and in particular Lagrange multipliers of general massive gravity theory. Our notation and conventions  are briefly explained in the Appendix.

\section{Constrained Variational Derivatives}
We find it convenient to start  our discussion in the next section with the  general definitions and  the remarks about the variational procedure we shall use throughout the paper. The field equations of Einstein's General Theory of Relativity can be derived from Einstein-Hilbert (EH) action
\begin{equation}
I_{tot}
=
\int_U R*1+I_m, 
\end{equation}
via a variational procedure, where the gravitational Lagrangian is given by $\mathcal{L}_{EH}=R*1$ and $R$ is the Ricci scalar constructed from the Riemann curvature tensor and $*1$ is the invariant volume element. The action for non-gravitational  fields coupling to gravity is denoted by $I_m$. It turns out that any theory which claims to generalize or modify Einstein's GR must include classically well-tested GR as a special case and reduce to GR (and hence Newtonian gravity) in the weak gravity regime \cite{c-will}.

In our work, we shall not address problems arising from certain modified gravitational Lagrangians or their coupling to matter or the phenomenological implications  of these theories but  scrutinize  methods of the derivation of field equations that follow from a given modified gravitational action.
The dimensionality of spacetime turns out to be a crucial parameter for the gravitational models, hence we shall consider some popular three dimensional cases after  we derive general formulae.

Relative to an orthonormal coframe,   metric variational derivatives   are induced by coframe variational derivatives and that the vanishing torsion condition can in fact be  implemented into the coframe variation either by introducing Lagrange multipliers  or directly incorporating this condition into the variation by using the variational derivative $\delta\Theta^\alpha=0$, which relates the variational derivatives of the connection and coframe 1-forms. Using  either  of  the methods,  one ends up with the same set of metric field equations. However, the latter scheme  gets increasingly cumbersome  with increasing complexity of  Lagrangians at hand. Therefore, we shall use outline the Lagrange multiplier method appropriate to our investigation.

In summary, we will work out the Riemannian content of the generalized/modified  gravitational Lagrangians which contain the  scalars  built out of the curvature tensor and its various contractions. Specifically, we will consider the gravitational Lagrangian densities  of type $\mathcal{L}^{(n)}_{G}=f(\mathcal{S})*1$ where $\mathcal{S}$ is any scalar built out of curvature  tensor and the function $f$ is an arbitrary differentiable function of its argument. In that, we shall assume that  $n$-form $\mathcal{S}*1$ is expressible in terms differential forms such as coframe and connection 1-forms, curvature 2-forms and their contraction using Hodge dual operator.  As will be apparent, for practical convenience, we adopt the Lagrange multiplier method  \cite{buchdahl,buchdahl2,kichenasssamy,kopczynki,der-tuc1,hehl-mccrea}
using differential forms language (summarized in the appendix for convenience)  adapted from \cite{thirring} throughout the paper. Although the Lagrange multiplier method can also be carried out relative to a coordinate coframe,  the use of differential forms enhances the practical use of the Lagrange multiplier method. Using the  method that we adopted to derive metric field equations for $f(R)$ theory, Palatini $f(R)$ equations can be derived from the same set of  variational derivatives, which can be considered as one of the advantages that makes the Lagrange multiplier is more efficient  method. We shall work in $n\geqslant3$ dimensions, and a  gravitational action $I_G$ has the general form
\be
I_G=
\int_{U}\mathcal{L}^{(n)}_{G}
=
\int_{U} f(\mathcal{S})*1
\ee
in a local chart $U\subset M$ of $n$ dimensional pseudo-Riemannian manifold where    $\mathcal{L}^{(n)}_{G}$ is Lagrangian density $n$-form which already includes the invariant volume form and $\mathcal{S}$ is a scalar which can be expressed in terms of curvature tensor.  The metric field equations that follow are chart independent provided that the basis coframe  and the connection 1-forms have the following transformation properties
\begin{align}
\theta^\alpha
&\mapsto
\tilde{\theta}^\alpha
=
\Lambda^{\alpha}_{\fant{a}\beta}{\theta}^\beta,
\\
\omega^\alpha_{\fant{a}\beta}
&\mapsto
\tilde{\omega}^\alpha_{\fant{a}\beta}
=
(\Lambda^{-1})^{\alpha}_{\fant{a}\mu}\omega^\mu_{\fant{a}\nu}\Lambda^{\nu}_{\fant{a}\beta}
+
(\Lambda^{-1})^{\alpha}_{\fant{a}\mu}d\Lambda^{\mu}_{\fant{a}\beta},
\end{align}
for overlapping charts, where the functions $\Lambda^{\alpha}_{\fant{a}\gamma}$ take the values in the Lorentz group. In our presentation, we shall  suppress the integral sign and use Lagrangian density $n$-form for convenience. We start with deriving a general equation for the variational derivatives of generalized gravitational actions. The basic idea simply is  to use product rule for the variational derivatives of the scalar fields and subsequently convert the variational derivatives of the scalars into variational derivatives of expressions involving forms which allows us to adopt orthonormal coframe. Relative to an orthonormal coframe, the total variational derivative  of  the $n-$form $\mathcal{L}^{(n)}_{G}$ can be calculated  as follows. The Leibnitz rule for  variational derivatives of the product of two scalars gives
\be
\delta(f*1)
=
(\delta f)*1+f\delta *1,
\ee
where $\delta$ denotes variational derivative. The variational derivative of  the function $f$ can be written  as $\delta f=f'\delta S$ where $f'=\frac{df}{d\mathcal{S}}$. Consequently, the  variation can be written as
\be\label{add1}
\delta\mathcal{L}^{(n)}_{G}
=
f'(\delta\mathcal{S})*1+f\delta *1.
\ee
Now, using  the product rule for the variation
\be\label{add2}
\delta(\mathcal{S}*1)
=
(\delta\mathcal{S})*1
+
\mathcal{S} \delta*1
\ee
for the first term on the right hand side of Eqn. (\ref{add2}), Eqn. (\ref{add1}) can be written as
\be\label{add3}
\delta\mathcal{L}^{(n)}_{G}
=
f'\delta(\mathcal{S}*1)+(f-f'\mathcal{S})\delta *1.
\ee
Note here that the second term on the right hand side of Eqn. (\ref{add3}) involves only the coframe variational terms whereas the variational derivatives of the gravitational variables, such as the connection 1-forms, come from the first term. The second term on the right hand side of Eqn. (\ref{add3}) can be simplified by recalling that relative to an orthonormal coframes $\theta^\alpha$, for $\alpha=0, 1, 2\cdots (n-1)$, the volume $n$-form can be written as
\be\label{add4}
*1
=
\tfrac{1}{n!}\varepsilon_{\alpha_1\alpha_2\alpha_3\cdots\alpha_n}\theta^{\alpha_1\alpha_2\alpha_3\cdots\alpha_n}
\ee
where $\varepsilon_{\alpha_1\alpha_2\alpha_3\cdots\alpha_n}$ completely antisymmetric permutation symbol. Using the expression (\ref{add4}), the variational derivative of the volume form takes the form
\begin{align}
\delta*1
&=
\tfrac{1}{n!}\varepsilon_{\alpha_1\alpha_2\alpha_3\cdots\alpha_n}
\left(\delta\theta^{\alpha_1}\wdg\theta^{\alpha_2\alpha_3\cdots\alpha_n}
+
\theta^{\alpha_1}\wdg\delta\theta^{\alpha_2}\wdg\theta^{\alpha_3\cdots\alpha_n}
+
\cdots
+
\theta^{\alpha_1\alpha_2\alpha_3\cdots\alpha_{(n-1)}}\wdg \delta\theta^{\alpha_n}\right)
\nonumber\\
&=
\delta\theta^{\alpha_1}\wdg\tfrac{1}{(n-1)!}\varepsilon_{\alpha_1\alpha_2\alpha_3\cdots\alpha_n}\theta^{\alpha_2\alpha_3\cdots\alpha_n}\label{add5}
\end{align}
Note that, unlike exterior derivative, the variational derivative $\delta$ does not pick up a minus sign passing through a coframe 1-form and that
$\delta$ does not commute with $*$.  Thus, relabeling the dummy indices $\alpha_1$  in (\ref{add5}) and the definition of Hodge dualilty for basis
coframe 1-forms, the expression for the coframe variation of the volume element in Eqn. (\ref{add5})  can be reduced to
\be\label{add6}
\delta*1
=
\delta\theta^{\alpha}\wdg*\theta_{\alpha}.
\ee
Finally, using Eqn. (\ref{add6}) and Eqn. (\ref{add3}), the  expression for the total variation of the action density $\mathcal{L}^{(n)}_{G}$
takes the following convenient form
\be\label{der2}
\delta\mathcal{L}^{(n)}_{G}
=
f'\delta(\mathcal{S}*1)+\delta\theta^\alpha \wdg (f-f'\mathcal{S})*\theta_\alpha.
\ee
This key equation can be regarded  as a kind of master formula which has a wide generality, as will be shown below.
It expresses the variational derivative of the expression $f(\mathcal{S})*1$ in terms of the simpler variational derivative $n$-form $\mathcal{S}*1$ plus additional terms. For instance, for the simplest case $\mathcal{S}=R$, which will be discussed in the next section, in which the Lagrangian  density becomes $\mathcal{L}^{(n)}_{GEH}=f(R)*1$   and the total variation will given by Eqn. (\ref{der1}), is the simplest example of the general expression (\ref{der2}). It is possible to consider slightly more general Lagrangian density where   $f$ is a function of a number scalars, say for example,  $\mathcal{L}^{(n)}_{G}=f(\mathcal{S}, \mathcal{R})*1$. In this case, the total variational derivative takes the form
\be\label{der3}
\delta\mathcal{L}^{(n)}_{G}
=
\frac{\partial f}{\partial\mathcal{S}}\delta(\mathcal{S}*1)+\frac{\partial f}{\partial\mathcal{R}}\delta(\mathcal{R}*1)
-
\delta\theta^\alpha \wdg \left(\frac{\partial f}{\partial\mathcal{S}}\mathcal{S}
+
\frac{\partial f}{\partial\mathcal{R}}\mathcal{R}-f\right)*\theta_\alpha,
\ee
where  $f$ is now assumed to be an analytical function of  two distinct curvature scalars $\mathcal{S}$ and $\mathcal{R}$.
Not to burden the presentation with unnecessary complications, we shall mainly consider the gravitational Lagrangian  of type $f(\mathcal{S})*1$.
In order to evaluate the total variational derivative of (\ref{der2}) with respect to independent connection and coframe 1-forms relative to
an orthonormal basis explicitly, the first term on the right hand side of (\ref{der2}) is to be expressed  in terms of coframe 1-forms, connection 1-forms, curvature 2-form and  contractions of curvature 2-forms. The second term on the right hand side of (\ref{der2}) directly contributes to the coframe variational derivative and hence to the metric field equations. In the Lagrange multiplier method we  shall use, the independent field variables  are the coframe and connection 1-forms as well as the Lagrange multiplier $(n-2)$-forms. However, the  connection 1-forms are constrained to be torsion free
using a Lagrange multiplier $(n-2)$ form in $n$-dimensional space. The  vanishing of the  the torsion cannot be imposed straightforwardly to the variation derivatives  since it is a dynamical constraint. In such a constrained theory, the coframe variations get contributions involving  Lagrange multiplier forms.

The equivalence of the Lagrange multiplier method with the metric variation relative to a coordinate coframe  has been shown in \cite{safko}. However, it  should be emphasized that the Palatini method  and  the metric method, except for the only the case of the Einstein-Hilbert action $\mathcal{L}^{(n)}_{EH}$,  lead to different field equations.

As remarked before, relative to an orthonormal coframe metric equations are induced by coframe variations and the equations obtained by independent connection variation can be used to solve the Lagrange multiplier $(n-2)$-forms in terms  of other fields. The Lagrange multiplier $n$-form terms, which contains vector-valued  Lagrange multiplier $(n-2)$-forms $\lambda^\alpha$, of the form
\be\label{multiplier-def}
\mathcal{L}_{LM}[\lambda_\alpha, \omega^{\alpha}_{\fant{a}\beta},\theta^\alpha]
=
\lambda_{\alpha}\wdg \Theta^\alpha
=
\lambda_{\alpha}\wdg (d\theta^\alpha+\omega^{\alpha}_{\fant{\mu}\beta}\wdg \theta^\beta),
\ee
and  $\lambda_\alpha$ is promoted to a new field  variable. The variation of the total action is $\mathcal{L}^{(n)}_{tot}=\mathcal{L}^{(n)}_{GEH}+\mathcal{L}_{LM}$  with respect to the independent variables $\theta^\alpha$, $\omega_{\alpha\beta}$
and $\lambda_\alpha$. The metric field equations of the original Lagrangian corresponds  to the metric field equations of a section of the $\mathcal{L}_{tot}$ determined by $\Theta^\alpha=0$ and the field equations obtained from connection variation allows one to express $\lambda_\alpha$ in terms of the other fields.  The total variation of the total Lagrangian density $\mathcal{L}^{(n)}_{tot}$  with respect to the independent variables with the torsion constrained to vanish has the general expression
\be\label{general-total-var}
\delta\mathcal{L}^{(n)}_{tot}
=
\delta\theta^{\alpha}\wdg
\Big(
\frac{\delta\mathcal{L}^{(n)}_{G}}{\delta\theta^{\alpha}}
+
D\lambda_\alpha
\Big)
+
\delta\omega_{\alpha\beta}
\wdg\left\{
\Pi^{\alpha\beta}-\tfrac{1}{2}(\theta^{\alpha} \wdg\lambda^\beta-\theta^\beta\wdg\lambda^{\alpha})
\right\}
+
\delta\lambda_\alpha\wdg \Theta^\alpha,
\ee
up to an omitted  total exterior derivative and the antisymmetric $(n-1)$-form $\Pi^{\alpha\beta}\equiv\delta\mathcal{L}^{(n)}_{G}/\delta\omega_{\alpha\beta}$ is defined for convenience. The metricity condition for the torsion free connection relative to an orthonormal coframe can be expressed in terms of connection 1-forms as $\omega_{\alpha\beta}+\omega_{\beta\alpha}=0$. Therefore, the  metricity constraint  can be implemented by anti-symmetrizing
the coefficient of $\delta\omega_{\alpha\beta}$ in expression for the total variation. However, a nonmetricity  constraint for connection can also be imposed by introducing another Lagrange multiplier $(n-1)$-form $\rho_{\alpha\beta}=\rho_{\beta\alpha}$ with the term  $\rho^{\alpha\beta}\wdg(\omega_{\alpha\beta}+\omega_{\beta\alpha})=\rho^{\alpha\beta}\wdg D\eta_{\alpha\beta}$ added to the original Lagrangian density. The Lagrange multipliers $\rho_{\alpha\beta}$ can be found from the symmetric part of the variational derivative with respect to the connection 1-form in this case. We will not investigate the modified  theories of gravity with torsion and non-metricity  (metric-affine gauge theories of gravity) \cite{hehl-mccrea}. In this work, we  simply assume that the connection is metric compatible and therefore it has the anti-symmetry property, $\omega_{\alpha\beta}+\omega_{\beta\alpha}=0$ and accordingly $\delta\omega_{\alpha\beta}+\delta\omega_{\beta\alpha}=0$ in all the total variations we consider. Consequently, one ends up with the property
$\Pi_{\alpha\beta}+\Pi_{\beta\alpha}=0$. There is an equivalence between  the antisymmetric $\Pi_{\alpha\beta}$  which  are tensor-valued $(n-1)$-forms and the Lagrange multiplier  $\lambda^\alpha$  which are vector-valued  $(n-1)$-forms  \cite{hehl-mccrea}. The field equations $\delta\mathcal{L}^{(n)}_{tot}/\delta\omega_{\alpha\beta}=0$ can  uniquely be solved for the  Lagrange multiplier $(n-2)$ forms as
\be\label{general-lag-mult}
 \lambda^\alpha
 =
2i_\beta\Pi^{\beta\alpha}
 +\tfrac{1}{2} \theta^\alpha\wedge i_\mu i_\nu\Pi^{\mu\nu},
\ee
by calculating two successive contractions of the equations $\delta\mathcal{L}^{(n)}_{tot}/\delta\omega_{\alpha\beta}=0$. For $n=3$, the wedge product in the second term (\ref{general-lag-mult}) is unnecessary since $\lambda^\alpha$ is 1-form in three dimensions. The resulting expression then can be used in the $D\lambda^\alpha$ term  for the metric equation ${\delta\mathcal{L}^{(n)}_{tot}}/{\delta\theta_{\alpha}}=0$. The trace of this term, subject to the constraint $\Theta^\alpha=0$, has the following general expression
\be\label{contraction-of-lag-term}
\theta_\alpha\wdg D\lambda^\alpha
=
2di_\beta(\theta_\alpha\wdg \Pi^{\alpha\beta}),
\ee
which applies to all the cases we consider. The equations $\delta\mathcal{L}^{(n)}_{tot}/\delta\lambda_{\mu}=\Theta^\mu=D\theta^\mu=0$ imply that the resulting covariant exterior derivative  $D$ is torsion-free and metric compatible, and therefore $D^2\theta^\alpha=\Omega^{\alpha}_{\phantom{a}\beta}\wdg\theta^\beta=0$.

In the following sections, we shall explicitly work out the Riemannian content of the field equations obtained from metric variational principle  for various   generalized/modified gravitational actions and we shall assume that zero-torsion and metricity conditions are imposed for all the models. In particular, we make  use  of the  general expressions (\ref{der2}) and (\ref{general-total-var})-(\ref{contraction-of-lag-term}). On the other hand we shall indicate how the field equations can be obtained adopting the  Palatini approach in comparison. If a matter field couples not only to metric components \cite{note-hodge} but also to the connection, the resulting theory  is known as Riemann-Cartan  theory where torsion is produced by spin density of the matter \cite{kopczynki}. However, as has been noted before, torsion may also be generated by gradient of  scalar field of a scalar-tensor  theory, such as BD theory in  which BD scalar field couples non-minimally to the curvature. On the basis of equivalence between $f(R)$ theory and the BD type scalar-tensor theory one also expects this to be the case for generic Palatini $f(R)$ theories. Although we shall not address the problem of formulating   Riemann-Cartan type theory for the gravitational Lagrangians   coupled with spinor matter  Lagrangians, our unified approach is even appropriate  to  accommodate such an extension.

\section{Some Previous Results: $f(R)$ Theory}\label{sect2}

As an illustration of the general method of calculation for variational derivative of  Lagrangian densities that we  shall present, we first derive the  field equations for the well-known $f(R)$ theory. In this section, we shall study both metric and Palatini  $f(R)$ theories in relation to respective scalar-tensor theories.

\subsection{Metric $f(R)$ Theory}
We start by recalling that, relative to an orthonormal coframe, the Einstein-Hilbert $n$-form Lagrangian  density can be written in the following equivalent forms
\be\label{EH-lag}
\mathcal{L}^{(n)}_{EH}
=
\tfrac{1}{2}\Omega_{\alpha\beta}\wdg *\theta^{\alpha\beta}=\tfrac{1}{2}R*1,
\ee
where $*1=\theta^{01\cdots (n-1)}$ is the oriented volume element, $R=i_\nu i_{\mu}\Omega^{\mu\nu}=i_{\nu}R^{\nu}$ is the scalar curvature and the dimension of the spacetime is assumed to  have the values $n\geqslant3$. From purely mathematical point of view, the Lagrangian density (\ref{EH-lag}) can be assumed to have the generalization of the form
\be\label{GEH-lag}
\mathcal{L}^{(n)}_{GEH}
=
f(R)*1,
\ee
where the function $f$ is assumed to be an arbitrary algebraic  scalar function that is differentiable with respect to its argument. Similar to the formulae given  in (\ref{der2}), the variational derivative  of the total Lagrangian  density $\delta\mathcal{L}^{(n)}_{GEH}$  can be written as
\be\label{der1}
\delta\mathcal{L}^{(n)}_{GEH}
=
f'\delta\mathcal{L}^{(n)}_{EH}-\delta\theta^\alpha \wdg (f'R-f)*\theta_\alpha,
\ee
where $f'=\frac{df}{dR}$.
The total variational derivative  of the total Lagrangian  density $\mathcal{L}_{tot}[\theta^\alpha, \omega^{\alpha}_{\fant{q}\beta}, \lambda_\alpha]=\mathcal{L}^{(n)}_{GEH}+\mathcal{L}_{LM}$ with respect to the independent  variables  can be written as
\begin{align}
\delta\mathcal{L}^{(n)}_{tot}
=&
\delta\omega_{\alpha\beta}\wdg \tfrac{1}{2}\left\{D(f'*\theta^{\alpha\beta})
-
(\theta^{\alpha}\wdg \lambda^{\beta}-\theta^{\beta}\wdg \lambda^{\alpha})\right\}
 \nonumber\\
 &+
 \delta\theta_\mu\wdg
\left\{
 f'\Omega_{\alpha\beta}\wdg*\theta^{\alpha\beta\mu}
+
(f-Rf') *\theta^\mu
+
D\lambda^\mu
 \right\}
 +\delta\lambda_{\alpha}\wdg\Theta^\alpha\label{geh-metric-var},
\end{align}
 up to a  disregarded closed form,  and from this expression one can read off
\be\label{pi-0}
\Pi^{\alpha\beta}
=
D(f'*\theta^{\alpha\beta}).
\ee
Using the general formula (\ref{general-lag-mult}) and the result (\ref{pi-0}), the Lagrange multiplier can be found as
\be\label{ray}
\lambda^\alpha
=
2*(df'\wdg\theta^\alpha).
 \ee
Consequently, using this expression for the Lagrange multiplier in  the equation obtained from the coframe  variations, one finds the metric field equations induced by the coframe variations  as
\be\label{0-torsion-metric-eqn}
-f'*G^{\alpha}
+
\tfrac{1}{2} (f-Rf')\wdg *\theta^\alpha
 +
D*(df'\wdg\theta^{\alpha})=0,
\ee
which was actually derived long ago \cite{buchdahl-first-fr}. Here $*G^\alpha=-\frac{1}{2}\Omega_{\mu\nu}\wdg *\theta^{\alpha\mu\nu}$ is Einstein $(n-1)-$ form.
An equation for the scalar function $f(R)$ can be found by tracing the metric equation (\ref{0-torsion-metric-eqn}). This gives
\be\label{trace-g}
(n-1)d*df'
=
\left(
\frac{n}{2}f-f'R
\right)*1,
\ee
which is identically satisfied for $f(R)=R$ for the vacuum case. Moreover, for the Einstein-Hilbert case where $f(R)=R$, the Lagrange multiplier vanishes identically since $\lambda^\alpha=0$. For this case, the Palatini formalism leads to the same metric field equations as in the metric formalism (see Eqns. (\ref{torsion-eqn-RC}) and  (\ref{metric-eqn-RC})) as well. We also note the special case $f(R)=R^2$ for which the trace equation (\ref{trace-g}) reduces to the homogeneous wave equation for scalar curvature $d*dR=0$ in four spacetime dimensions $n=4$. However, $R^2*1$ term survive for  $n=3$ and for $n>4$.
As can be observed from the field equations presented below, the spacetime dimension is usually a  crucial parameter that may determine physical features of the model at hand. Note that, as expected on consistency grounds, for $f(R)=R$, Eqns. (\ref{0-torsion-metric-eqn}) yield the Einstein vacuum field equations. However, unlike Einstein field equations, they contain 4$^{th}$ order partial derivatives of the metric components relative to the coordinate coframe because of the presence of the term $D*(df'\wdg\theta^{\mu})$. This term drops in the constant curvature case $\Omega^{\mu\nu}=k\theta^{\mu\nu}$. In this case, scalar curvature is constant, $R=kn(n-1)$ and $df'=0$ identically. Thus, it is possible to find a Lagrangian density $f(R)$ for which isotropic spaces are  vacuum solutions.
Inserting $\Omega^{\mu\nu}=k\theta^{\mu\nu}$ into (\ref{0-torsion-metric-eqn}), one finds $f(R)=e^{k(n-1)R}$. Cosmological dynamics based on
the exponential gravitational  Lagrangian is  studied in \cite{abdelwahab} and in generic $f(R)$ models $R=0$ does not obviously  imply $R=0$ as in GR.
In particular, we note here that, for another simple  case $f(R)=R^m$ which we shall refer to in the succeeding sections, Eqn. (\ref{0-torsion-metric-eqn}) yields
\be\label{f(R)-metric-eqn}
-
R*G^{\alpha}
-
\frac{(m-1)}{2m}R^2*\theta^\alpha
 +
\tfrac{1}{2} (m-1)(m-2)dR\wdg i^{\alpha}*dR
+
\tfrac{1}{2}(m-1)
D*(dR\wdg\theta^{\alpha})
=0,
\ee
which holds for $n\geqslant3$ spacetime dimensions.  The solution to the Einstein vacuum field equations $R^{\alpha\beta}=0$  trivially solves  the field equations derived from Lagrangian density $f(R)=R^m$. Thus, for example, static spherically symmetric  solutions to the Einstein vacuum field equations
are also solutions to (\ref{f(R)-metric-eqn}). But the converse is not necessarily true. Not all the vacuum solutions to $f(R)$ theory
are also solutions to  the vacuum Einstein field equations.

\subsection{Palatini/Riemann-Cartan Type $f(R)$ Theory}

Riemann-Cartan type gravitational theories  accommodate torsion, which is assumed to be generated by spin \cite{kerlick}. However, in addition to spin, scalar-tensor theories also give rise to torsion in terms of gradient of a scalar field \cite{dereli-tucker-spin-bd}. As we shall show below, for generic $f(R)$ theory, the theory admits an algebraic torsion and consequently  the Riemann-Cartan connection can be expressed in terms of Riemannian connection and a contorsion 1-form. In this case, the metric field equations can be cast into a form completely expressed in terms of Riemannian geometrical quantities. In the  Palatini method  to calculate the variational derivative to $f(R)$ theory, the field equations for connection is used to solve  the connection   in terms of Riemannian connection and additional terms resulting from algebraic torsion in terms of the gradient of the scalar $df'$.  The Palatini approach to $f(R)$
theory, which we shall  study in this section, is therefore can be considered as modified gravitational theory with algebraic torsion. We  shall start with formally the same Lagrangian density (\ref{EH-lag}). In the case where non-zero torsion is allowed \cite{sotir,capozzi},
the field equations,  which can be found from the variational derivative (\ref{geh-metric-var}), are
\begin{align}
df'\wdg*\theta^{\alpha\beta}+f'\Theta^{\mu}\wdg*\theta^{\alpha\beta}_{\fant{qa}\mu}
&=0,
\label{torsion-eqn-RC}\\
f'*G^\alpha
+
(f-f'R) *\theta^\alpha
&=0,
 \label{metric-eqn-RC}
\end{align}
respectively.  Note here that for $f(R)=R$, the Palatini variation yields $\Pi^{\alpha\beta}=0$ identically if one imposes $\Theta^{\alpha}=0$ after the variation in the case of $\mathcal{L}^{(n)}_{EH}$. This follows from the fact that  in the metric variation for the Einstein-Hilbert Lagrangian, one finds from (\ref{ray}) that  $\lambda^\alpha=0$ \cite{ray}. As a result Palatini and metric variational derivatives lead to the identical metric equations  in this exceptional case of  the Einstein-Hilbert Lagrangian. The field equation which follows from the independent connection variation,  Eqn. (\ref{torsion-eqn-RC}) is an algebraic equation for the torsion 2-form $\Theta^\alpha$ which can be solved for it as
\be
\Theta^\alpha
=
\tfrac{1}{(n-2)}\theta^\alpha\wdg d\ln f',
\ee
using the identity $\theta^\alpha\wdg i_\alpha\lambda=p\lambda$ which holds for arbitrary $p$-form $\lambda$. This result then can be used to solve the corresponding connection 1-forms which can be obtained from the Cartan's first structure equations.

Returning to the Palatini  equations,   the variational derivative with respect to connection, namely $\Pi^{\alpha\beta}=0$ lead to an algebraic
torsion in terms of the gradient $df'$. As a result, in Palatini formalism,  even if the connection is metric compatible (and therefore commutes with
contractions), it has torsion. However, algebraic torsion allows the elimination of the independent connection 1-form and the field equations can be cast
into a metric  theory with  additional gravitational currents under suitable conditions stated below. For this purpose, it is convenient to introduce contorsion 1-form $K_{\alpha\beta}=-K_{\beta\alpha}$ defined by $\Theta^\alpha=K^{\alpha}_{\fant{a}\beta}\wdg \theta^\beta.$ The contorsion 1-forms can be expressed in term of the exterior derivatives of the function $f'$ as
\be\label{contort-fr}
K^{\alpha\beta}
=
\tfrac{1}{(n-2)}\left\{\theta^{\alpha}(d\ln f')^\beta-\theta^{\beta}(d\ln f')^\alpha\right\},
\ee
where the indices of the 1-form $d\ln f'$ are relative to the orthonormal coframe we employ. In terms of contorsion 1-form, the metric compatible connection 1-form $\dot{\omega}^{\alpha}_{\fant{a}\beta}$ can be written as sum of Riemannian connection  ${\omega}^{\alpha}_{\fant{a}\beta}$ and a contorsion term
as
\be\label{connection-contort-decomp}
{\omega}^{\alpha}_{\fant{a}\beta}
=
\dot{\omega}^{\alpha}_{\fant{a}\beta}+K^{\alpha}_{\fant{a}\beta},
\ee
where, in order to distinguish metric connection from the Riemannian connection, the dotted symbols for the Riemannian quantities are used.
The  result  (\ref{connection-contort-decomp}) will be elaborated below.

The Cartan's second structure equations for the corresponding curvature 2-form then lead to similar decomposition of the curvature 2-forms
as well. One finds
\be
{\Omega}^{\alpha\beta}
=
\dot{\Omega}^{\alpha\beta}
+
\dot{D}K^{\alpha\beta}+K^{\alpha}_{\fant{a}\mu}\wdg K^{\mu\beta},
\ee
where $\dot{\Omega}^{\alpha\beta}$ is the Riemannian curvature 2-form and $\dot{D}$ is the  covariant exterior derivative corresponding to Riemannian covariant derivative. Note that $f$ and hence $f'$ is a function of scalar curvature ${R}$. Thus, the elimination of independent connection requires ${R}$ to be expressed in terms of the dotted  quantities  using the trace of the metric equations  (\ref{metric-eqn-RC}). If we assume that this requirement can be satisfied, then the Einstein $(n-1)$-form can be written as
\be
f'*{G}^\alpha
=
f' *\dot{G}^\alpha
-
\dot{D}*(df' \wdg \theta^\alpha)
-
\frac{(n-1)}{(n-2)}
\frac{1}{f' }
*T^\alpha[f' ],
\ee
where the energy momentum forms for the scalar field $f'$, namely
\be\label{scalar-fr-en-mom}
*T_\alpha[f' ]
=
-
\tfrac{1}{2}\left\{(i_\alpha df' )*df'
+
df' \wdg i_\alpha*df'
\right\},
\ee
is defined for convenience. Note that this result explicitly shows that non-zero torsion produces the kinetic term for the function $f'$
at the level of action and this corresponds to the effective energy momentum  term  $*T^\mu[f' ]$ in the resulting metric equation. Finally,
using this result and dropping the dots over Riemannian quantities, the metric field equations can be written as
\be\label{metric-eqn-reduced-RC}
f' *{G}^\alpha
=
\tfrac{1}{2} (f-Rf')*\theta^\alpha
+
{D}*(df' \wdg \theta^\alpha)
+
\frac{(n-1)}{(n-2)}
\frac{1}{f' }
*T^\alpha[f' ],
\ee
which are expressed in terms of Riemannian connection 1-form and the corresponding Riemannian  curvature tensor.
Taking the trace of the metric field equation  one finds
\be
(n-1)d*df'
=
\tfrac{1}{2}(n-1)\frac{1}{f'}df'\wdg*df'+\tfrac{1}{2}f'R*1
\ee
Note that because of the term $D*(df'\wdg\theta^{\alpha})$, the field equations (\ref{metric-eqn-reduced-RC}) are fourth order partial differential equations relative to a coordinate basis. The above reduction of $f(R)$ theory with torsion to a metric theory has been carried out recently using the equivalence of such a theory with Scalar-Tensor theories relative to a coordinate basis  and it is also argued that the theories with torsion and/or non-metricity can also be cast
into metric theory for $f(R)$ theories \cite{sotir}. As we shall show explicitly in the following sections,  this is not true
for other modified gravitational Lagrangians.

We now briefly discuss the equivalence of generic $f(R)$ theory with torsion to the Scalar-tensor theory. Note that if  the field variable $f'(R)$ is promoted to a scalar field,   then equations of motion (\ref{0-torsion-metric-eqn}) and (\ref{trace-g}) for $f(R)$ theory have formal resemblance to the metric equations in BD theory except for energy momentum  $(n-1)$-form corresponding to  the kinetic term for the BD scalar function.  In other words, the equations of motion for generic  $f(R)$ theory can be obtained from a Brans-Dicke type action with kinetic term for BD scalar that minimally couples to the metric tensor is absent. Consider the following BD-type Lagrangian density in $n$ dimensions
\be\label{bd-action}
\mathcal{L}
=
\mathcal{L}_{BD}
+
V(\phi)*1
=
-
\frac{\phi}{2}\Omega_{\mu\nu}\wdg*\theta^{\mu\nu}
+
\frac{\omega}{2\phi}d\phi\wdg*d\phi
+
V(\phi)*1,
\ee
where the scalar field $\phi$ couples to the metric non-minimally and there is a self interaction potential term for the scalar field. The metric field
equations, that follow from (\ref{bd-action}), is
\be\label{bd-metric-eqn}
\phi*G^\alpha
=
-\frac{\omega}{\phi}*T^\alpha[\phi]
+
D*(d\phi\wdg \theta^\alpha)
+
V(\phi)*\theta^\alpha,
\ee
where the canonical  energy momentum $(n-1)$-form field $*T^\alpha[\phi]$, which results from  commuting the coframe variation with Hodge dual, is
\be\label{scalar-en-mom}
*T_\alpha[\phi]
=
-\tfrac{1}{2}\{(i_\alpha d\phi)*d\phi+d\phi\wdg i_\alpha*d\phi\}
\ee
whereas the field equation for scalar  field can be written as
\be\label{bd-scalar}
\left(\omega+\tfrac{n-1}{n-2}\right)d*d\phi
=
V'*1.
\ee
For the BD theory with vanishing BD-parameter, namely for $\omega=0$, the metric and the scalar field  equations that follow from the variation of the action for this Lagrangian with zero torsion condition imposed can be written as
\begin{align}
-&\phi*G^\alpha
+
D*(d\phi\wdg \theta^\alpha)
+
V(\phi)*\theta^\alpha=0,\label{res1-metric}
\\
&R
=V'(\phi)\label{res1-scalar},
\end{align}
respectively \cite{ohan}. Here, $'$ denotes derivative with respect to the argument of the scalar potential, $V'=\frac{dV}{d\phi}$. The trace of the metric equation (\ref{res1-metric}) and (\ref{res1-scalar}) can be combined to have
\be\label{res2-scalar}
(n-1)d*d\phi
=
\left\{nV(\phi)-\tfrac{1}{2}(n-2)\phi V'(\phi)\right\}*1.
\ee
Comparison with the metric field equations for the $f(R)$ theory, assuming that $f$ is a definite functional form and that $f''\neq0$, the equation
$f'(R)=\phi$ can be inverted in order to be express $R$ in terms of $\phi$. Then, $R(\phi)$ can be used in the equation
\be
nV(\phi)-\tfrac{1}{2}(n-2)\phi V'(\phi)
=
\tfrac{n}{2}f-f'R ,
\ee
to determine  the scalar potential $V(\phi)$. In particular, for $n=4$ one has $V(\phi)=\tfrac{1}{2}f(R(\phi))$. As a result, with the above identification and redefinition of the field variables, the field equations for $f(R)$ theory transformed into a BD-type theory. In the BD-type scalar theory with field Eqns. (\ref{res1-metric}) and (\ref{res1-scalar}), the energy momentum $(n-1)$-forms for any matter field $*T^\alpha_m[\Phi]$ that couples only to the metric field but not to the derivatives of the metric field, are covariantly constant. Explicitly, using the contracted second Bianchi identity, which is a purely geometrical identity, namely $D*G^\alpha=0$, the covariant derivative of (\ref{res1-metric}) gives
\be
-d\phi\wdg*G^\alpha
+
\Omega^{\alpha}_{\fant{a}\beta}\wdg *(d\phi\wdg \theta^\beta)+V'(\phi)d\phi\wdg*\theta^\alpha
=
0,
\ee
by virtue of the field equation (\ref{res1-scalar}). In other words, $D*T^\alpha_m[\Phi]=0$ follows from the fact that the covariant exterior derivative
of the vacuum  metric field equations (\ref{res1-metric}) vanish. Conversely, the metric field equations (\ref{res1-metric}) yield the field equation
(\ref{res2-scalar}) by taking the covariant exterior derivative of (\ref{res1-metric}) and using (\ref{res1-scalar}). The fact that the field equations are covariantly constant can be shown to follow from more general property of Lagrangian density that it has diffeomorphism invariance, independent of the resulting field equations.

It is important to emphasize again that, the Palatini and the metric variational principles in general, lead to different sets of metric field equations. The scalar tensor equivalence for the corresponding $f(R)$ theories above provides an example for the inequivalence of the two theories since the Palatini variation $f(R)$ theory lead BD-type scalar tensor theory with BD parameter $\omega=-\tfrac{(n-1)}{(n-2)}$ with potential term for the scalar field whereas $f(R)$ theory based on metric variational principle lead to scalar-tensor theory with $\omega=0$
\cite{sotiriou-palatini,sotir}.

The Palatini connection is conformally related to Levi-Civita as   in (\ref{connection-contort-decomp}) with the contorsion form expressed in terms of  a scalar gradient (\ref{contort-fr}). Because the conformal factor involves scalar curvature, the parallel transport of the connection  in the Palatini method leads to a distinct autoparallel curves which can be expressed in terms of Levi-Civita connection.  A timelike vector field $X$, which is normalized as $g(X,X)=-1$, is tangent  to autoparallel curves of the connection (\ref{connection-contort-decomp}) if it satisfies \be\label{autoparallel-eqn}
\nabla_XX=0.
\ee
Here, it is convenient to introduce the operator $\flat$ which maps the vectors to their metric duals, i.e. to the corresponding 1-forms. It can be defined implicitly in terms of the  metric tensor as $X^\flat(-)=g(X,-)$. Since  $\nabla_X$ is metric compatible and hence commutes with the operator $\flat$,
 we have $\nabla_XX^\flat=(\nabla_XX)^\flat$. Using the expression (\ref{connection-contort-decomp}) for the connection 1-form, Eqn. (\ref{autoparallel-eqn})
can be written in terms of the Riemannian covariant derivative $\dot{\nabla}_X$ as
\be\label{geodesic-eqn2}
\dot{\nabla}_XX^\flat+\frac{f''}{2f'}i_X(dR\wdg X^\flat)=0.
\ee
As a result, similar to the scalar tensor theories \cite{dereli-tucker-autoparallel}, the non-Riemannian connection given in (\ref{connection-contort-decomp})
induces   an acceleration term depending on the gradient of the scalar curvature into the autoparallel curves. We note here that the geodesics are determined by the Riemannian structure instead of the  independent Palatini connection as a result of the fact that the  matter energy momentum $(n-1)$-forms are covariantly  constant: $D*T^\alpha_m=0$ \cite{tomi-koivisto}.  On the other hand, the correct cosmological behavior of $f(R)$ theories in the Palatini  formalism requires  consistent  averaging procedure of energy momentum forms  describing the  microscopic(atomic) structure of matter fields and it also predicts alterations to atomic structure and particle physics laws, see e.g. \cite{mota}.

\subsection{Equivalence to Scalar-tensor Theories at the Level of Action }

The equivalence of $f(R)$ theory to particular Scalar-tensor theories discussed above by comparing the corresponding field equations can also be deduced, perhaps in an indirect way, at the level of action. (Such an  equivalence holds for $f(R)$ theory with/without torsion.) The equivalence  can be achieved by introducing  an auxiliary scalar field $\chi$ and a scalar Lagrange multiplier $\lambda$ imposing the constraint $\chi-\mathcal{S}=0$ to the  gravitational action.

We shall mainly consider modified gravitational Lagrangians  of the form $f(\mathcal{S})*1$, where $f$ is an algebraic function of scalar $\mathcal{S}$
and we assume that $S*1$ can be expressible in terms of basis coframe forms, connection 1-forms,  curvature 2-forms and their various contractions.
 In the place of $f(\mathcal{S})*1$, consider an auxiliary  Lagrangian density $\mathcal{L}_{aux}=\mathcal{L}_{aux}[\chi,\lambda, \theta^\alpha, \omega^{\alpha}_{\fant{q}\beta}]$  of the form
\be\label{aux-action}
\mathcal{L}_{aux}
=
[f(\chi)
+
\lambda (\chi-\mathcal{S})]*1.
\ee
The total variational derivative  of the auxiliary Lagrangian (\ref{aux-action}) with respect to the independent variables indicated above yields
\be\label{aux-action-tot-var}
\delta\mathcal{L}_{aux}
=
[\delta\chi(f'+\lambda)
+
\delta\lambda(\chi-\mathcal{S})]*1
-
\lambda\delta (\mathcal{S}*1)+\delta\theta^{\alpha}\wdg(f+\lambda\mathcal{S})*\theta_{\alpha}
\ee
where  $\delta$ denotes the variational derivative as before and the prime denotes derivative with respect to the auxiliary field $\chi$.
The  equation for the scalar field $\chi$ immediately yield the Lagrange multiplier  $\lambda=-f'$.  Replacing this result and
the corresponding constraint  back into the action (\ref{aux-action}) one finds the dynamically equivalent Lagrangian density as
\be\label{gen-reduced-action1}
 \mathcal{L}_{aux-eqv.}
=
[f(\chi)
+
(\chi-\mathcal{S})f'(\chi)]*1,
 \ee
up to an overall sign. The resulting  Lagrangian density $ \mathcal{L}_{aux-eqv.}$ now has no dependence on  the Lagrange multiplier
$\lambda$ and in the equivalent Lagrangian density $f'$ couples to the scalar  $\mathcal{S}$. However, the reduced form of the auxiliary
Lagrangian density allows redefinition  of the field variables. For $f(R)$ theory, namely for $\mathcal{S}=R$, Eqns. (\ref{aux-action-tot-var})
and (\ref{gen-reduced-action1}) reduce to
\begin{align}
\delta\mathcal{L}_{aux}
&=
[\delta\chi(f'+\lambda)
+
\delta\lambda(\chi-R)]*1
-
\lambda\delta (R*1)
-
\delta\omega_{\mu\nu}\wdg D(\lambda*\theta^{\mu\nu})
\nonumber\\
&
+\delta\theta^{\alpha}\wdg\left\{-\lambda\Omega_{\mu\nu}\wdg*\theta^{\mu\nu}_{\fant{aa}\alpha}+(f+\lambda R)*\theta_{\alpha}\right\},
\end{align}
and
\be\label{aux-f(r)}
 \mathcal{L}_{aux-eqv.}
=
f'\Omega_{\alpha\beta}\wdg*\theta^{\alpha\beta}
+
(f-f'R)*1,
\ee
respectively. Although the auxiliary action (\ref{aux-f(r)}) consistently simplifies to the original action, this form of the action  further suggests the redefinition  $f'(R)=\phi$. This allows one to bring the action (\ref{aux-f(r)}) into equivalent Scalar-tensor type action given in (\ref{bd-action}) with $\omega=0$ and  with an appropriate potential term for the scalar field assuming that $\phi$ can be written in terms of $R$. This general  mechanism of generating non-minimally coupled scalar field in generalizing  gravitational actions will be used to propose a modified   Chern-Simons   theory in four dimensions  in a later section.

\section{Variational Derivatives of Modified Gravitational Actions}

In addition to the generic $f(R)$ theories, modified gravitational  theories based on the Lagrangian densities involving quadratic curvature invariants  and their algebraic generalizations also lead to fourth order models. These models have a long history \cite{schimming-schmidt}, they have been studied in different
contemporary contexts ranging from cosmology and perturbative approach to quantum theory gravity \cite{stelle}. This section is devoted to the calculation of the metric field equations for fourth and sixth order theories.

\subsection{Generalized Gauss-Bonnet Actions}\label{sect3}

The outcome of employing  extra dimensions for unification as well as quantum corrections of gravity requires some well behaving generalizations of the EH Lagrangian, such as scalars constructed from of higher order curvature terms. Actually, using an arbitrary scalar in action constructed from Riemann tensor leads to pathologies \cite{zwiebach,ghosts} such as to have more than required degrees of freedom in equations of motion or existence of ghosts. However, there is a special combination of such terms in action, as shown by Lovelock \cite{Lovelock}, which are known as dimensionally continued Euler forms or Lovelock scalars. It turns out that, the Lagrangians constructed from these terms have some desirable properties such as, though nonlinearly, the field equations involve at most second derivatives of metric components \cite{Lovelock,lanczos}, theory is free of ghosts \cite{zwiebach,zumino}, i.e. kinetic terms in the action with the wrong sign, and in four dimensions the only surviving term apart from cosmological constant is the EH Lagrangian \cite{Lovelock,bach}. This generalization of considering Euler forms were studied in different contexts \cite{madore,mullerhoissen,Choquet-Bruhat,deruelle,arikdereli,deruellemadore} such as Kaluza-Klein reduction, low energy limit of string theory, brane-worlds, and higher dimensional exact solutions.

Thus, in higher dimensions, one can consider, in addition to the usual Einstein-Hilbert term, the combination of Euler forms
\be
I_G
=
\sum^{[n/2]}_{k=0}\mathcal{L}^{(k,n)}_{L},
\ee
where
\be
\mathcal{L}^{(k,n)}_{L}
=
\Omega_{\alpha_1\beta_1}\wdg\cdots\wdg\Omega_{\alpha_k\beta_k}\wdg*\theta^{\alpha_1\beta_1\cdots\alpha_k\beta_k}.
\ee
For $k=2n$, $\mathcal{L}^{(n,n)}_{L}$ is an exact form and is proportional to Euler-Poincare characteristic of even dimensional manifold. For simplicity,  we consider the gravitational Lagrangian density $n$-form
\begin{align}
\mathcal{L}^{(n)}_{\mathcal{G}}
&=
\Omega_{\alpha\beta}\wdg\Omega_{\mu\nu}\wdg*\theta^{\alpha\beta\mu\nu}
\nonumber\\
&=
2\Omega_{\alpha\beta}\wdg*\Omega^{\alpha\beta}
-
4R_\alpha\wdg*R^\alpha+R^2*1\label{GB-lag},
\end{align}
corresponding to $k=2$ for $n\geqslant 4$. This term is called as Gauss-Bonnet or Lanczos Lagrangian, it is a pure divergence in $d=4$, and first introduced by Lanczos \cite{lanczos}. Although here we only consider Gauss-Bonnet term, it is straightforward to extend the formalism to the  higher order Euler form Lagrangian densities. Now, it is convenient to define  the scalar  $\mathcal{G}=R_{\alpha\beta\mu\nu}R^{\alpha\beta\mu\nu}-4R_{\mu\nu}R^{\mu\nu}+R^2$ which  separates  the volume form so that $\mathcal{L}^{(n)}_{\mathcal{G}}=\mathcal{G}*1$. In four dimensions, $\mathcal{L}^{(4)}_{\mathcal{G}}$ is a total differential and does not contribute to the field equations. This  follows from the identity
\be\label{lanczos}
\tfrac{1}{2}\varepsilon^{\alpha\beta}_{\fant{\alpha\beta}\mu\nu}\Omega^{\mu\nu}
=
\Omega^{\alpha\beta}
-
\tfrac{1}{2}(\theta^\alpha\wdg R^\beta-\theta^\beta\wdg R^\alpha)
+
\tfrac{1}{4}R\theta^{\alpha\beta},
\ee
which relates the left dual of the curvature 2-form  to the curvature 2-form itself and its contractions \cite{benn-tucker}.
From (\ref{lanczos}), it follows that
\be
\Omega_{\alpha\beta}\wdg\Omega_{\mu\nu}\wdg*\theta^{\alpha\beta\mu\nu}
=
-\tfrac{1}{2}\Omega^{\alpha}_{\fant{a}\beta}\wdg\Omega^{\beta}_{\fant{a}\alpha},
\ee
as a result of first Bianchi identity and the right hand side is an exact form (see Eqn. (\ref{curv-tr})). Therefore,  the GB term of the form $\Omega_{\alpha\beta}\wdg\Omega_{\mu\nu}\wdg*\theta^{\alpha\beta\mu\nu}$ does not  contribute to the field equations. However, it is possible to make $\mathcal{L}^{(4)}_{\mathcal{G}}$ to contribute to field equations, for example, by coupling it with a scalar field. On the other  hand, the presentation of $f(R)$ type Lagrangian densities indicates  that $f(\mathcal{G})*1$  type Lagrangian density will also have similar features that $f'$  will act like a non-minimally coupled scalar field. After these preliminary remarks, we will derive the equations of motion for the modified Lagrangian density of the form
\be\label{GGB-lag}
\mathcal{L}^{(n)}_{G\mathcal{G}}
=
f(\mathcal{G})*1,
\ee
for $n\geqslant4$.
The independent variations with respect to the coframe and  connection 1-forms are constrained to be torsion free by Lagrange multiplier term  (\ref{multiplier-def}). Using the same steps in the derivation of the variation (\ref{der2}) from the modified Lagrangian (\ref{GEH-lag}), the total variation of the Lagrangian density,
\be
\mathcal{L}^{(n)}_{tot}
=
\mathcal{L}^{(n)}_{G\mathcal{G}}+\mathcal{L}_{LM},
\ee
is related to the variational derivative of the Lagrangian density $\mathcal{L}^{(n)}_{\mathcal{G}}$ as
\be\label{var-gb-gen-ver}
\delta\mathcal{L}^{(n)}_{tot}
=
f'\delta\mathcal{L}^{(n)}_{\mathcal{G}}-\delta\theta^\alpha \wdg(f'\mathcal{G}-f)*\theta_\alpha+\delta\mathcal{L}_{LM}.
\ee
Note that for $n=4$, there is no contribution from the term containing $\delta\mathcal{L}^{(n)}_{\mathcal{G}}$ on the RHS of (\ref{var-gb-gen-ver}). Thus, the cases $n>4$ and $n=4$ lead to different  field equations. The variations on the right hand side of this expression can be calculated to find
\begin{align}
\delta\mathcal{L}^{(n)}_{tot}
&=
\delta\omega_{\alpha\beta}
\wdg\left\{
\Pi^{\alpha\beta}-\tfrac{1}{2}(\theta^{\alpha} \wdg\lambda^\beta-\theta^\beta\wdg\lambda^{\alpha})
\right\}
+
\delta\lambda_\alpha\wdg \Theta^\alpha
\nonumber\\
&\phantom{=}+
\delta\theta_{\gamma}\wdg
\{
f'\Omega_{\mu\nu}\wdg\Omega_{\alpha\beta}\wdg i^\gamma*\theta^{\mu\nu\alpha\beta}
+
(f'\mathcal{G}-f)*\theta^\gamma
+
D\lambda^\gamma
\}.
\end{align}
Here $'$ denotes derivatives with respect to the scalar $\mathcal{G}$ and the expression for $\Pi_{\alpha\beta}$ obtained from  the variational derivative with respect to  the connection 1-form yields
\be
\Pi^{\alpha\beta}
=
2D\left(f'\Omega_{\mu\nu}\wdg *\theta^{\mu\nu\alpha\beta}\right),
\ee
which is to be evaluated subject to the vanishing torsion condition. In contrast to the other field equations that we shall consider below, the field equations for the GB terms are second order in metric components. This follows from the fact that the second Bianchi identity satisfied by the curvature 2-forms and the  Lagrange multiplier terms can be expressed in terms of  curvature 2-forms. The order of the metric equations persist even in the corresponding first order formalism where  an independent  dynamical torsion is present.

As a result, the expressions for the Lagrange multipliers can be inserted into  the metric equations
$\delta\mathcal{L}^{(n)}_{tot}/\delta\theta_{\lambda}=0$ to find
 \begin{equation}
 f'\Omega_{\mu\nu}\wdg\Omega_{\alpha\beta}\wdg *\theta^{\mu\nu\alpha\beta\lambda}
+
(f-f'\mathcal{G})*\theta^\lambda
-
4\Omega_{\mu\nu}\wdg D*(df'\wdg\theta^{\mu\nu\lambda})
=0,\label{gb-n-greater-4}
\end{equation}
for $n>4$ dimensions. The field equations (\ref{gb-n-greater-4})  that  follow from the  $f(\mathcal{G})$ Lagrangian are similar to the field equations for the Lagrangians  with a   scalar field which couple non-minimally to the GB-terms. The term $\Omega_{\mu\nu}\wdg D*(df'\wdg\theta^{\mu\nu\lambda})$ contains 4$^{th}$ order partial derivatives of the metric components relative to a coordinate coframe. Now specializing to $n=4$ dimensions, in particular, to the gravitational Lagrangian $\mathcal{L}^{(4)}_{tot}=\mathcal{L}^{(4)}_{EH}+\mathcal{L}^{(4)}_{GGB}$, the metric field equations can be written as
\be
-*G^\alpha
+
(f-f'\mathcal{G})*\theta^\alpha
-
4\Omega_{\mu\nu}\wdg D*(df'\wdg\theta^{\mu\nu\alpha})=0.
\ee
Apparently, the order of the metric field equations for $f(\mathcal{G})$ augmented to  four as a result of algebraic generalization of GB term.
We refer to the review article \cite{felice} for $f(\mathcal{G}$  models and their applications.

Before we begin to study the  the individual terms appearing in (\ref{GB-lag}), we note that the dimension of spacetime is crucial and in particular,
in four spacetime dimensions, using (\ref{GB-lag}) and (\ref{lanczos}) the most general Lagrangian  which are quadratic in the scalars  the curvature  can be written in the form
\be\label{des-bay}
\mathcal{L}^{(4)}_{G}
=
a R^\alpha\wdg *R_\alpha+b R^2*1,
\ee
with $a,b$ being  constants. The  Lagrangian density (\ref{des-bay}) is  the most general gravitational Lagrangian density  involving quadratic curvature terms in three spacetime  dimensions as well and we will study of this special case in a later section below. However, the most general Lagrangian involving quadratic curvature terms  is to include $R_{\alpha\beta\mu\nu}R^{\alpha\beta\mu\nu}$ term in $n>4$ dimensions.

\subsection{Ricci-Squared Action}

In this subsection we will calculate  the variational derivatives of the Ricci-squared Lagrangian density whereby the corresponding field equations for  the gravitational Lagrangian densities in (\ref{des-bay}) with arbitrary $a, b$ and its algebraic generalizations of the form $f(R, \mathcal{P})$ can be found using the results of the previous sections. We first consider a Lagrangian density containing Ricci tensor-squared term, which can be written in terms of Ricci 1-forms as follows
\be
\mathcal{L}^{(n)}_{\mathcal{P}}
=
R_{\alpha}\wdg*R^{\alpha}
=
R_{\alpha\beta}R^{\alpha\beta}*1.
\ee
The metric field  equations corresponding to   $\mathcal{L}^{(n)}_{\mathcal{P}}$ will also be used later in the study of three dimensional massive gravity models. We first derive field equations  for the Lagrangian density    $\mathcal{L}^{(n)}_{\mathcal{P}}$ in a form appropriate to our presentation and then derive the field equations for its generalization of the form $\mathcal{L}^{(n)}_{G\mathcal{P}}=f(\mathcal{P})*1$ where $\mathcal{P}\equiv R_{\alpha\beta}R^{\alpha\beta}$.
 We first note that the symmetry of the Ricci tensor indices $R^\alpha\wdg \theta_\alpha=0$ follows from the contraction of the first Bianchi identity, namely
$D\Theta^\alpha=\Omega^{\alpha}_{\fant{a}\beta}\wdg\theta^\beta=0$  together with the anti-symmetry property $\Omega^{}_{\alpha\beta}+\Omega^{}_{\beta\alpha}=0$ of the curvature 2-forms relative to an orthonormal coframe. Thus, the symmetry of the components of the Ricci tensor can be incorporated conveniently into the total variation without introducing an additional constraint. We constrain the variation to be torsion-free so that we have the total Lagrangian density
$
\mathcal{L}^{(n)}_{tot}
=
\mathcal{L}^{(n)}_{\mathcal{P}}
+
\mathcal{L}_{LM}
$.
This Lagrangian has   the total variation
\begin{align}
\mathcal{L}^{(n)}_{tot}
&=
\delta\omega_{\alpha\beta}\wdg
\left\{
D*(\theta^\alpha\wdg R^\beta-\theta^\beta\wdg R^\alpha)
-
\tfrac{1}{2}
(\theta^\alpha\wdg \lambda^\beta-\theta^\beta\wdg \lambda^\alpha)
\right\}
+
\delta\lambda_\alpha\wdg \Theta^\alpha
\nonumber\\
&+
\delta\theta^\mu
\wdg
\left\{
\Omega_{\alpha\beta}\wdg i_\mu*(\theta^\alpha\wdg R^\beta-\theta^\beta\wdg R^\alpha)
-
i_\mu(R_\alpha\wdg*R^\alpha)
+
D\lambda_\mu
\right\}\label{ricci-var},
\end{align}
where the identity $R_{\mu\nu}*1=\Omega^{}_{\alpha\mu}\wdg*\theta^{\alpha}_{\fant{a}\nu}$ has been used to obtain (\ref{ricci-var}). The resulting vacuum metric field equations, subject to the condition that $\Theta^\alpha=0$, can be conveniently written in terms of  1-form $\mathcal{P}^\alpha=\mathcal{P}^{\alpha}_{\fant{a}\beta}\theta^\beta$ as
$
*\mathcal{P}^{\alpha}
=0
$
where $*\mathcal{P}^{\alpha}$ is given by
\be\label{ricci-form2}
*\mathcal{P}^\mu
=
\Omega_{\alpha\beta}\wdg i^\mu*(\theta^{\alpha}\wdg R^\beta-\theta^{\beta}\wdg R^\alpha)
-
i^\mu(R_\alpha\wdg*R^\alpha)
+
2Di_\alpha D *(\theta^\alpha\wdg R^\mu-\theta^\mu\wdg R^\alpha).
\ee
To our knowledge, the metric field equation for the Ricci-squared model formulated in terms of differential forms (\ref{ricci-form2}) is new. The covariant exterior derivatives in the Lagrange multiplier  terms explicitly show that the Ricci-squared model is a fourth order model. 1-form $\mathcal{P}_\mu$ defined by (\ref{ricci-form2}) is symmetric with respect to its indices $\mathcal{P}_{\alpha\beta}=\mathcal{P}_{\beta\alpha}$. We also note here that the trace of the metric field equations (\ref{ricci-form2}) can be found as
\be\label{ricci-trace}
\mathcal{P}^{\alpha}_{\fant{a}\alpha}*1
=
(n-4)R^\alpha\wdg*R_\alpha
+
\tfrac{n}{2}d*dR.
\ee
The trace formula (\ref{ricci-trace}) again reflects the fact that the spacetime dimension is important for the Ricci-squared models as we shall see in three dimensions.  This form of the field equations allows one to find a nontrivial solution to  the $\Lambda$-vacuum field equations almost by inspection of the metric field equations. For example, $R^\alpha=k\theta^\alpha$  with the constant $k$ solves the $\Lambda-$vacuum field equations $*\mathcal{P}^\alpha=-2(n-1)k^2*\theta^\alpha$.

Now, having the total variational derivative  of   $\mathcal{L}^{(n)}_{\mathcal{P}}$ at our disposal, the field equations for $\mathcal{L}^{(n)}_{G\mathcal{P}}=f(\mathcal{P})*1$ can be found. As before, we also assume  that $f$ is an algebraic and differentiable function of its argument. The corresponding metric field equations  can be found  by straightforward  application of (\ref{der2}). The total variation of
 $\mathcal{L}^{(n)}_{tot}=\mathcal{L}^{(n)}_{G\mathcal{P}}+\delta\mathcal{L}^{}_{LM}$ then becomes
 \be
\delta\mathcal{L}^{(n)}_{tot}
=
f'\delta\mathcal{L}^{(n)}_{\mathcal{P}}-\delta\theta^\alpha \wdg (f'\mathcal{P}-f)*\theta_\alpha
+
\delta\mathcal{L}^{}_{LM}.
\ee
Carrying out the total variation  using (\ref{der2}) one can  find
 \begin{align}
\delta\mathcal{L}^{(n)}_{tot}
&=
\delta\omega_{\alpha\beta}
\wdg\left\{
D*f'(R^\alpha\wdg \theta^\beta-R^\beta\wdg \theta^\alpha)
-
\tfrac{1}{2}
(\theta^\alpha\wdg \lambda^\beta-\theta^\beta\wdg \lambda^\alpha)
\right\}
+
\delta\lambda_\alpha\wdg \Theta^\alpha
\nonumber\\
&\phantom{=}+
\delta\theta^{\mu}
\wdg
\left\{
2f'
\Omega_{\alpha\beta}\wdg i_\mu*(R^\alpha\wdg \theta^\beta)
-
f'i_{\mu}(R_\alpha\wdg*R^\alpha)
+
(f'\mathcal{P}-f)*\theta_\mu
+
D\lambda_\mu
\right\}.
\end{align}
For the purpose of comparison, it is convenient to write the  resulting metric field equation using 1-form  $\mathcal{P}_\alpha$ defined in (\ref{ricci-form2}). This gives
 \be\label{ricci-pal}
 f'*\mathcal{P}^\alpha+D(f'\lambda^\alpha)+(f'\mathcal{P}-f)*\theta^\alpha=0.
 \ee
In contrast to the metric formalism, if the  Palatini formalism is  used as in \cite{baojiu-barrow-mota}, then from (\ref{ricci-var}), the equation for the connection  can be found as
\be\label{ricci-palatini}
\Pi^{\alpha\beta}=D*f'(\theta^{\alpha}\wdg R^\beta-\theta^{\alpha}\wdg R^\beta)=0,
\ee
where $D$ in this case is non-Riemannian covariant exterior derivative. Eqn. (\ref{ricci-palatini})  has to be solved for connection 1-forms and then subsequently replaced into the coframe variation to obtain the metric field equations. Obviously, as in the case of $f(R)$ theory, it follows from (\ref{ricci-palatini})  that the Palatini and metric formalism lead to totally different set of equations for the metric field. However, adopting Palatini formalism with an independent and \emph{torsion-free} connection, the recent work on the models based on the Lagrangian $n$-form    of the generic form $f(\mathcal{P})*1$ lead to the result that such models accommodate metrics of Einstein spaces \cite{francaviglia}.

Returning to our treatment of the models with quadratic curvature invariants using the metric formalism, we now study   another popular fourth order model with the Lagrangian density containing Riemann tensor-squared (Krectschmann scalar) terms  \cite{yang}, namely the $n$-form
\be
\mathcal{L}^{(n)}_{\mathcal{R}}
=
\tfrac{1}{2}
\Omega_{\alpha\beta}\wdg*\Omega^{\alpha\beta}
=
\tfrac{1}{4}R_{\alpha\beta\mu\nu}R^{\alpha\beta\mu\nu}*1,
\ee
and its algebraic generalization is of the form
\be
\mathcal{L}^{(n)}_{G\mathcal{K}}
=
f(\mathcal{K})*1,
\ee
with $f$ being an algebraic function of the Krectschmann  scalar $\mathcal{K}=R_{\alpha\beta\mu\nu}R^{\alpha\beta\mu\nu}$. The field equations for  this Lagrangian, with torsion tensor constrained to vanish, can be obtained from the variation of the total Lagrangian density
\be
\mathcal{L}^{(n)}_{G\mathcal{K}}
=
\mathcal{L}^{(n)}_{\mathcal{K}}+\mathcal{L}_{LM}.
\ee
For the Lagrangian density $\mathcal{L}^{(n)}_{G\mathcal{R}}$,  equation (\ref{der2}) becomes
\be
\delta\mathcal{L}^{(n)}_{G\mathcal{K}}
=
f'\delta\mathcal{L}^{(n)}_{\mathcal{K}}-\delta\theta^\alpha \wdg \tfrac{1}{2}(f'\mathcal{K}-f)*\theta_\alpha
+
\delta\mathcal{L}^{}_{LM}.
\ee
The total variation with respect to independent  coframe  and connection 1-forms and Lagrange multiplier  $(n-2)$-forms then can be written as
\begin{align}
\delta\mathcal{L}^{(n)}_{tot}
&=
\delta\omega_{\alpha\beta}
\wdg\left\{
D*(f'\Omega^{\alpha\beta})+\tfrac{1}{2}(\lambda^{\alpha}\wdg \theta^\beta-\lambda^\beta\wdg \theta^{\alpha})
\right\}
+
\delta\lambda_\alpha\wdg \Theta^\alpha
\nonumber\\
&\phantom{=}+
\delta\theta^{\mu}\wdg
\left\{
-
\tfrac{1}{2}f'\left(
i_\mu\Omega_{\alpha\beta}\wdg*\Omega^{\alpha\beta}
-
\Omega_{\alpha\beta}\wdg i_\mu*\Omega^{\alpha\beta}\right)
+
\tfrac{1}{2}(f'\mathcal{K}-f)*\theta_\mu
+
D\lambda_\mu
\right\}.
\end{align}
The field equations $\delta\mathcal{L}^{(n)}_{tot}/\delta\omega_{\alpha\beta}=0$ can be solved for the Lagrange multipliers $\lambda^\alpha$ using (\ref{general-lag-mult}) with $\Pi^{\alpha\beta}=D(f'*\Omega^{\alpha\beta})$. Therefore, the metric field equations subject to the condition that $\Theta^\alpha=0$, can be written as
\be\label{gen-sky-eqn}
4Di_\alpha D(f'*\Omega^{\alpha}_{\fant{a}\mu})
-
\theta_\mu\wdg Di_{\alpha}i_{\beta}D(f'*\Omega^{\alpha\beta})
+
f'*T_\alpha[\Omega]
+
\tfrac{1}{2}(f'\mathcal{K}-f)*\theta_\mu
=0,
\ee
where the canonical energy momentum  tensor for the curvature 2-forms,
\be\label{en-mom-curv-2form}
*T_\mu[\Omega]
=
-\tfrac{1}{2}
\left(
i_\mu\Omega_{\alpha\beta}\wdg*\Omega^{\alpha\beta}
-
\Omega_{\alpha\beta}\wdg i_\mu*\Omega^{\alpha\beta}\right),
\ee
are introduced  for convenience. Note that this metric equation holds in $n\geqslant 3$ dimensions. For $f*1=\Omega_{\alpha\beta}\wdg *\Omega^{\alpha\beta}$, and $f'=1$, the field equations (\ref{gen-sky-eqn}) reduce to
\be\label{sky-eqn}
Di_\alpha D*\Omega^{\alpha\mu}
-
\tfrac{1}{4}\theta^\mu\wdg Di_{\alpha}i_{\beta}D*\Omega^{\alpha\beta}
+
*T^\mu[\Omega]
=0,
\ee
which we shall make use of in the following sections. The trace of Eqns. (\ref{sky-eqn}) leads to the following equation
\be\label{sky-trace}
(n-1)d*dR+(n-4)\Omega^{\alpha\beta}\wdg*\Omega_{\alpha\beta}=0,
\ee
using  the Bianchi identity $D*G^\alpha=0$. As for  Ricci-squared and $f(R)=R^2$ models, Eqn. (\ref{sky-trace}) depends of the dimensions of spacetime where $n=4$  appears to be special.

The method adopted can be used to calculate the field equations easily for more complicated-looking actions compared to coordinate methods. As an illustration of this point consider  generalized gravitational Lagrangian density in  curvature scalar, for example,  of the form $f(\mathcal{G})=\mathcal{G}^2$,
\be
\mathcal{G}^2
=
\mathcal{K}^2+16\mathcal{P}^2+R^4-8R^2\mathcal{P}-8\mathcal{K}\mathcal{P}+2\mathcal{K}R^2,
\ee
contains cross terms of the form $R^2\mathcal{K}$, $R^2\mathcal{K}$ and $\mathcal{K}\mathcal{P}$. The variational derivatives  of these types of individual cross terms  as well as the models based on the generalized Lagrangian of type, for example $f(R,\mathcal{G})$, are of some interest. In order to illustrate how the method we adopted can easily be used to calculate the variational derivatives  of Lagrangian densities in curvature polynomials, consider the generalized Lagrangian density of the form
\be
\mathcal{L}^{(n)}_{R^2\mathcal{K}}
=
R^2R^{\alpha\beta\mu\nu}R_{\alpha\beta\mu\nu}*1.
\ee
As before we constrain the  variation by adding $\mathcal{L}_{LM}$ to the corresponding  action and perform a total variation
\begin{align}
\delta\mathcal{L}^{(n)}_{R^2\mathcal{K}}
&=
2R\mathcal{K}\delta\mathcal{L}^{(n)}_{EH}
+
R^2\delta(\Omega_{\alpha\beta}\wdg*\Omega^{\alpha\beta})+\delta\mathcal{L}_{LM}.
\end{align}
This explicitly gives
\begin{align}
\delta\mathcal{L}^{(n)}_{R^2\mathcal{K}}
&=
\delta\omega_{\alpha\beta}
\wdg\left\{
D\left\{(\mathcal{K}R-R^2)*\theta^{\alpha\beta}\right\}
-
\tfrac{1}{2}(\theta^{\alpha}\wdg \lambda^\beta-\theta^\beta\wdg \lambda^{\alpha})
\right\}
+
\delta\lambda_\alpha\wdg \Theta^\alpha
\nonumber\\
&\phantom{=}+
\delta\theta^{\mu}\wdg
\left\{
-
\mathcal{K}R*G_\mu
-
R^2\left(
i_\mu\Omega_{\alpha\beta}\wdg*\Omega^{\alpha\beta}
-
\Omega_{\alpha\beta}\wdg i_\mu*\Omega^{\alpha\beta}\right)
+
D\lambda_\mu
\right\}.
\end{align}
Solving $\delta\mathcal{L}^{(n)}_{R^2\mathcal{R}}/\delta\omega_{\alpha\beta}=0$ for the Lagrange multipliers $\lambda_\alpha$ in terms of other field variables using (\ref{general-lag-mult}), and using the resulting expression for $\lambda_\mu$ in the equation $\delta\mathcal{L}^{(n)}_{R^2\mathcal{R}}/\delta\theta^{\mu}=0$, the metric  field equation becomes
\be\label{metric-eqn-cross-term-lag}
-
\mathcal{K}R*G^\alpha
+
2R^2*T^\alpha[\Omega]
+
D*\{d(R\mathcal{K}-R^2)\wdg \theta^\alpha\}=0.
\ee
As for all the generalized gravitational Lagrangians, written in the above form,  field equations has formal resemblance to the field equations  (\ref{bd-metric-eqn}) and (\ref{bd-scalar}) of BD type Scalar-tensor theory where scalar field is replaced by a function  of curvature scalars.

\subsection{Weyl Tensor-squared Action }

Weyl tensor is the trace-free part of the Riemann curvature tensor and therefore the corresponding Weyl 2-form $C_{\alpha\beta}$ satisfy $C_{\alpha\beta}\wdg*\theta^{\alpha\beta}=0$. Recall that relative to an orthonormal coframe in $n\geqslant4$ dimensions curvature  2-form has the expression \cite{thirring}
\be\label{weyl-def}
C^{\alpha\beta}
=
\Omega^{\alpha\beta}-\tfrac{1}{n-2}\left(\theta^{\alpha}\wedge R^{\beta}-\theta^{\beta}\wedge R^{\alpha}
\right)
+
\tfrac{1}{(n-1)(n-2)}R\, \theta^{\alpha\beta}.
\ee

It is conformally invariant,  has the same symmetry properties with the Riemann tensor and exists in spacetime dimensions $n\geqslant4$.  There is a unique  action based on the contraction of the Weyl 2-form with itself
\begin{align}
\mathcal{L}^{(n)}_W
&=
C_{\alpha\beta}\wdg *C^{\alpha\beta}
\nonumber\\
&=
\Omega_{\alpha\beta}\wdg *\Omega^{\alpha\beta}
-
\tfrac{2}{n-2}R^{\alpha}\wdg *R_\alpha
+
\tfrac{1}{(n-1)(n-2)}R^2*1.\label{BW-lag}
\end{align}
In four dimensional case, Lagrangian density (\ref{BW-lag}) is locally scale invariant and the metric field equations derived from $\mathcal{L}^{(4)}_W$ are
known as Bach-Weyl equations which have  4$^{th}$ order partial derivatives of the metric tensor relative to a coordinate basis \cite{bach}.

As we have mentioned above, the GB Lagrangian density  $\mathcal{L}^{(4)}_{\mathcal{G}}$ (\ref{GB-lag}) is a total differential. Thus, without altering the equations of motion for $\mathcal{L}^{(4)}_{W}$, by adding the  term $\mathcal{L}^{(4)}_{GB}$ to the Lagrangian density $\mathcal{L}^{(4)}_{W}$,  the term $\Omega_{\alpha\beta}\wdg *\Omega^{\alpha\beta}$ in the Lagrangian density $\mathcal{L}^{(4)}_{W}$  can be eliminated. This gives somewhat simplified four dimensional gravitational Lagrangian density
\be
\mathcal{L}^{(4)}_W
=
R^{\alpha}\wdg *R_\alpha
-
\tfrac{1}{3}R^2*1,
\ee
up to a total differential and an overall multiplicative constant. This reduction of the Lagrangian density makes it possible to derive the field equations of the model  $\mathcal{L}^{(4)}_{W}$ from those of of $f(R)=R^2$ and the Ricci-squared model studied  above. This requires the use of the properties and the identities satisfied by the  curvature 2-form and therefore the cases of the three  and higher dimensions has to be handled separately. Returning to the general ($n\geqslant4$) case  and following the preceding examples of generalized Lagrangian densities, it is convenient to define 0-form  $\mathcal{C}=C_{\mu\nu\alpha\beta}C^{\mu\nu\alpha\beta}$ and similarly the generalization of $\mathcal{L}^{(n)}_W$ is assumed to have the form
\be
\mathcal{L}^{(n)}_{GW}
=
f(\mathcal{C})*1,
\ee
where $f$ is an algebraic function of the scalar $\mathcal{C}$. Note that the local scale  invariance of the original Lagrangian  has been lost in such a generalization. We will study the metric field equations with torsion constrained to vanish and consider the total Lagrangian density $\mathcal{L}^{(n)}_{tot}=\mathcal{L}^{(n)}_{GW}+\mathcal{L}_{LM}$. The variational derivative, in the light of the variational relation given (\ref{der2}), is then given by
\be
\delta\mathcal{L}^{(n)}_{tot}
=
f'\delta\mathcal{L}^{(n)}_{W}+\delta\theta^\alpha \wdg(f-f'\mathcal{C} )*\theta_\alpha+\delta\mathcal{L}_{LM},
\ee
where $'$ in this case denotes derivative with respect to the scalar $\mathcal{C}$. Carrying out the variations indicated in this equations one finds
\begin{align}
\delta\mathcal{L}^{(n)}_{tot}
&=
\delta\theta^{\alpha}\wdg
\left\{
\frac{4f'}{n-2}R_\beta\wdg *C^{\beta\alpha}
+
f'*T^\alpha[W]
+
(f-f'\mathcal{C})*\theta^\alpha
+
D\lambda_\alpha
\right\}
\nonumber\\
&\qquad+
\delta\omega_{\alpha\beta}\wdg
\left\{
2D (f'*C^{\alpha\beta})
-
\tfrac{1}{2}(\theta^\alpha\wdg \lambda^{\beta}-\theta^\beta\wdg\lambda^{\alpha})
\right\}
+
\delta\lambda_\alpha\wdg \Theta^\alpha,
\end{align}
where the definition of canonical energy momentum tensor for the Weyl 2-form
\be
*T_\alpha[W]
=
-(i_\alpha C_{\mu\nu})\wdg *C^{\mu\nu}+C_{\mu\nu}\wdg i_\alpha  *C^{\mu\nu},
\ee
has been used for convenience and the irrelevant total exterior derivatives are omitted. The field equations $\delta\mathcal{L}^{(n)}_{tot}/\delta\omega_{\alpha\beta}=0$ can be solved for the Lagrange multipliers $\lambda^\alpha$ using (\ref{general-lag-mult}) with $\Pi^{\alpha\beta}=D(f'*C^{\alpha\beta})$. Therefore, the metric field equations subject to the condition that the torsion vanishes, becomes
\begin{align}
&f'\left\{\tfrac{4}{(n-2)}R_\beta\wdg *C^{\beta\alpha}
+
*T^\alpha[W]
+
4Di_\beta D*C^{\beta\alpha}
+
\theta^\alpha\wdg Di_{\mu}i_{\nu}D*C^{\mu\nu}\right\}
\nonumber\\
&+
(f-f'\mathcal{C})*\theta^\alpha
+
D i_\beta(df'\wdg* C^{\beta\alpha})
=0.\label{generalized-bach}
\end{align}
As a result, the  effect of algebraic generalization of the Weyl tensor-squared  action is  (i) the overall $f'$ factor for the first term and (ii) the last two improvement terms in (\ref{generalized-bach}). As expected for consistency of the field equations,  for $f(\mathcal{C})=\mathcal{C}$ the field equation\textbf{s} (\ref{generalized-bach}) reduce to Bach-Weyl equation \cite{bach,dereli-tucker-weyl}. As before, the Palatini variational derivative  also lead to
field equations that accommodate torsional connection for Weyl-squared action \cite{dereli-tucker-weyl}. Recently, in $n>4$ dimensions, the corrections to Tangherlini solution \cite{tangherlini} arising from higher order curvature terms, namely the terms of the form $\mathcal{C}*1$ are studied in \cite{frolov-shapiro} and in this context, it is possible to consider corrections originating   from  the terms of the form $f(\mathcal{C})*1$ in four dimensions   since  in four dimensions the Schwarzschild solution  is not modified by lowest quadratic curvature corrections of the form $R_\alpha\wdg *R^\alpha$ or $R^2*1$ \cite{frolov-shapiro}.

\subsection{Actions Involving Derivatives of Curvature Tensor}

In this section we  briefly discuss the gravitational Lagrangians involving derivatives of curvature invariants of type, for example  $R\Delta R$ \cite{buchdahl3}, where $\Delta=dd^\dagger+d^\dagger d$ is the Laplace-Beltrami operator. The codifferential  $d^\dagger$ is the metric dual of the exterior derivative $d$ and we shall use the definition   $d*=(-1)^{p+1}*d^\dagger$ acting on $p$-forms. The models based on this type of Lagrangians  as well as all of their algebraic generalizations considered above lead to the metric field equations  which are 6$^{th}$ order partial derivatives of metric components relative to  a natural coframe. The field equations for such theories involving  derivatives of scalar curvature  terms have found applications in cosmology \cite{hj-schmidt-var,gottlober}.

The calculations below can be extended to the generalized Lagrangians  of type $dR_{\alpha\beta}\wdg *dR^{\alpha\beta}$ and $dR_{\alpha\beta\mu\nu}\wdg *dR^{\alpha\beta\mu\nu}$ and also to their polynomial generalizations in a straightforward manner.

 We shall consider the simplest possible gravitational Lagrangian density involving derivatives of the curvature tensor,  namely $R\Delta R$ \cite{buchdahl3} and its algebraic generalizations. In order to apply the  variational procedure used above for generalized gravitational actions presented, it is first convenient to rewrite the action density $R\Delta R*1$ in the following form
\be\label{dR-Lagrange}
\mathcal{L}^{(n)}_{\partial R}
=
\tfrac{1}{2}dR\wdg*dR
=
-\tfrac{1}{2}(\partial R)^2*1,
\ee
equivalent to the expression $R\Delta R*1$ up to an omitted  total derivative and an overall sign. We define the scalar $(\partial R)^2$ which takes the form $ (\partial R)^2=g^{\alpha\beta}(\partial_{\alpha}R)(\partial_{\beta}R)$ relative to a coordinate frame. This definition  will be convenient in the subsequent generalization of the Lagrangian (\ref{dR-Lagrange}) below. The derivation of higher order field equations for gravitational Lagrangians of the form $R\Delta^m R*1$ is presented in \cite{hj-schmidt-var}.

As before, we constrain the total variation so that the torsion vanishes by adding $\mathcal{L}_{LM}$ to the original Lagrangian density
(\ref{dR-Lagrange}) as
\be
\mathcal{L}^{(n)}_{tot}
=
\tfrac{1}{2}dR\wdg*dR+\mathcal{L}_{LM}.
\ee
We found the total variational derivative  of the total  Lagrangian density to be
\begin{align}
\delta\mathcal{L}^{(n)}_{tot}
&=
\delta\omega_{\alpha\beta}\wdg
\left\{
2D ((\Delta R)*\theta^{\alpha\beta})
-
\tfrac{1}{2}(\theta^\alpha\wdg \lambda^{\beta}-\theta^\beta\wdg\lambda^{\alpha})
\right\}
+
\delta\lambda_\alpha\wdg \Theta^\alpha
\nonumber\\
&\qquad+
\delta\theta_{\alpha}\wdg
\left\{
(\Delta R) \Omega_{\mu\nu}\wdg*\theta^{\mu\nu\alpha}
-
R\Delta R*\theta^\alpha
+
*T^\alpha[R]
+
D\lambda^\alpha
\right\}\label{total-var-dr},
\end{align}
up to a closed form and the  canonical  energy momentum $(n-1)$-forms for the scalar curvature $R$ energy momentum forms of a scalar field,
\be
*T_\alpha[R]
=
-\tfrac{1}{2}\{(i_\alpha dR)*dR+dR\wdg i_\alpha*dR\},
\ee
has been used for convenience.  The canonical energy momentum form of the scalar curvature  results from the commutation of  variational derivative of the basis coframe 1-form with the Hodge dual exactly as in the case of scalar field (cf. Eqn (\ref{scalar-en-mom})). The resulting metric field equations \cite{buchdahl3, gottlober}, subject to the condition that the torsion vanishes,
are given by
\be\label{dR-metric eqn}
(\Delta R) *R^{\alpha}
+
\tfrac{1}{2}*T^\alpha[R]
-
D*\{(\Delta dR)\wdg\theta^\alpha\}=0.
\ee
These equations are  6$^{th}$ order in the components of metric tensor since $\lambda_\alpha$ turns out to have terms which are   5$^{th}$ order in the partial derivatives of the metric components. The trace of the metric field equations is given by
\be\label{dR-trace-eqn}
*(\Delta +R)\Delta R
-
\tfrac{(n-2)}{4(n-1)}
dR\wdg *d R
=
0.
\ee
Both (\ref{dR-metric eqn}) and (\ref{dR-trace-eqn}) have also formal resemblance to  field equations  of a BD type scalar-tensor theory (\ref{bd-metric-eqn}) and (\ref{bd-scalar}) respectively. If matter fields couple minimally to gravitational action (\ref{dR-Lagrange}), then the ``scalar field" $\Delta R$  act as a variable coupling  constant for the metric tensor. In this case, the trace of matter energy momentum fields acts as a source term for the wave equation  for the scalar $\Delta R$ in (\ref{dR-trace-eqn}).

As in the case for the $f(R)$ theory studied above,  for the sixth order theory based on (\ref{dR-Lagrange}), the Palatini variational derivative    lead to a metric  field equation different then (\ref{dR-metric eqn}) as well as it induces a torsion for the connection. In this case,  the connection 1-form can be found by solving the relation
\be
D((\Delta R)*\theta^{\alpha\beta})=0,
\ee
which lead to nonvanishing, algebraic torsion similar to the case for Riemann-Cartan  type  $f(R)$ theory.

Finally, we will consider the algebraic generalization of the sixth order theory based on (\ref{dR-Lagrange}).  The model then  has the Lagrangian density of the form
\be\label{dR-gen-lag}
\mathcal{L}^{(n)}_{G\partial R}
=
f((\partial R)^2)*1,
\ee
where, as in all the cases before, $f$ is assumed to be an algebraic function of the scalar $(\partial R)^2$.
The total variation of the total Lagrangian density
\be
\mathcal{L}^{(n)}_{tot}
=
\mathcal{L}^{(n)}_{G\partial R}+\mathcal{L}^{}_{LM},
\ee
with the $\Theta^\alpha=0$ condition imposed by Lagrange multiplier term can be found using (\ref{der2}). This gives
\be\label{gen-dr-var-exp1}
\delta\mathcal{L}^{(n)}_{tot}
=
f'\delta\mathcal{L}^{(n)}_{\partial R}-\delta\theta_{\alpha}\wdg \{(\partial R)^2 f'-f\}*\theta^{\alpha}+\delta\mathcal{L}^{}_{LM}.
\ee
Using the total variation (\ref{total-var-dr}), the field equations that follow  from the expression (\ref{gen-dr-var-exp1}) is
\be\label{fdR-metric eqn}
f'(\Delta R) *R^{\alpha}
-
\tfrac{1}{2}f'*T^\alpha[R]
-
D*\{f'(\Delta dR)\wdg\theta^\alpha\}
+
\tfrac{1}{2}\left\{(\partial R)^2f'-f\right\}*\theta^\alpha=0,
\ee
which is similar to the field equations based on the model (\ref{dR-Lagrange}) except for the the last terms in curly brackets and  the new factor $f'$ as the coefficient of the Ricci 1-form. $'$ denotes derivative with respect to $(\partial R)^2$. The generalized metric equations (\ref{fdR-metric eqn}) are  6$^{th}$ order in metric components which is of the same order  as the  model based on (\ref{dR-Lagrange}). It is interesting to note that, in contrast to the fourth order models, the model derived from the generalized gravitational Lagrangian density $\mathcal{L}^{(4)}_G=\mathcal{L}^{(4)}_{EH}+aR^2*1+bdR\wdg*dR$ is studied in \cite{gottlober} and is shown to be conformally equivalent to the conventional  General Relativity  coupled to two interacting scalar fields.

\section{Gravitational Chern-Simons Term in Four Dimensions}

The generalized gravitational actions studied in previous sections  indicate that  any algebraic function of curvature invariants at the level of action behaves like non-minimally coupled scalar field in the corresponding field equations. In this section, in the light of this important observation, we  find the field equations for gravitational action based on the generalization of gravitational Chern-Simons terms \cite{hehl-kopczynski,jackiw-pi, yunes}.
In $4n$ dimensions the Chern-Simons form in terms of 1-form-valued matrix notation can be written as
\be\label{curv-tr}
Tr[\Omega^{2n}]
\equiv
\Omega^{\alpha}_{\fant{q}\beta}\wdg\Omega^{\beta}_{\fant{q}\mu}\wedge\cdots\wdg\Omega^{\sigma}_{\fant{q}\nu}\wdg\Omega^{\nu}_{\fant{q}\alpha}
=
dK,
\ee
where $Tr$ denotes  trace and    $(4n-1)$-form $K$ can be written in terms of connection 1-form as
\be
K
=
2n\int^{1}_{0}dtt^{(2n-1)}Tr[\omega(d\omega+t\omega^2)^{(2n-1)}],
\ee
as a result of  the fact that $Tr[\omega^{4n}]=0$ \cite{chern-simons}. The indices and the wedge products are suppressed for simplicity of the notation. We  specialize to four dimensions, namely $n=1$, and in particular to the Einstein-Hilbert gravitational Lagrangian density modified by parity-violating Pontryagin density,  namely $\Omega_{\alpha\beta}\wdg \Omega^{\alpha\beta}$. It is well-known that it is an exact differential, namely,
\be
Tr[\Omega^2]
=
\Omega_{\alpha\beta}\wdg \Omega^{\alpha\beta}
=
d(\omega_{\alpha\beta}\wdg d\omega^{\alpha\beta}+\tfrac{2}{3}\omega_{\alpha\mu}\wdg \omega^{\mu}_{\fant{a}\nu}\wdg\omega^{\nu\alpha}),
\ee
and therefore it does  not yield any  field equations as it stands. However, in  the model of Jackiw-Pi \cite{jackiw-pi}, which is based on the Lagrangian density
 \be\label{original-JP-lag}
 \mathcal{L}^{(4)}_{JP}
 =
 \tfrac{1}{2}\Omega_{\alpha\beta}\wdg*\theta^{\alpha\beta}+\theta\,\Omega_{\alpha\beta}\wdg \Omega^{\alpha\beta},
 \ee
where the  $\Omega_{\alpha\beta}\wdg \Omega^{\alpha\beta}$  term couples to a dynamical scalar field $\theta$,  it does contribute to the Einstein field equations.  The scalar field $\theta$ can be regarded as Lagrange multiplier 0-form imposing  the constraint $\Omega_{\alpha\beta}\wdg \Omega^{\alpha\beta}=0$ on the resulting metric field equations,
\be\label{JP}
-*G^\alpha
+
4Di_\beta (d\theta\wdg \Omega^{\beta\alpha})
-
\theta^\alpha\wdg Di_\mu i_\nu (d\theta\wdg \Omega^{\mu\nu})=0,
\ee
which can be found by using (\ref{general-total-var}) and (\ref{general-lag-mult}). Since the gravitational Chern-Simons term does not contain Hodge dual operator, it does not contribute to the coframe field equation directly but  through Lagrange multiplier term in (\ref{general-total-var}).  Compared to Lagrange multiplier terms in  the field equations for actions containing quadratic curvature invariants studied above, the resulting metric field equations are 3$^{rd}$ order partial differential equations for the Chern-Simons modified theory. Using the general formula (\ref{contraction-of-lag-term}) together with  the first Bianchi identity and $\Omega_{\alpha\beta}+\Omega_{\beta\alpha}=0$, it is possible to show that neither of the Lagrange multiplier terms
on the right hand side of (\ref{JP}) has non-zero trace and therefore the vacuum field equations require $R=0$.

First order theory based on the Chern-Simons modified gravity (\ref{original-JP-lag}), with independent coframe and connection 1-forms has also been studied recently and leads to  dynamical torsion \cite{cantcheff}.

In the context  of the generalization of  gravitational actions we consider, instead of coupling  to a non geometrical  so-called cosmological scalar field $\theta$ it is possible to introduce a term of type $f(\mathcal{T})*1$ where  $*\mathcal{T}=\Omega_{\alpha\beta}\wdg \Omega^{\alpha\beta}$ is defined  for convenience. Therefore, within the same framework used for the other modified models above, it is natural to consider the Chern-Simons modified Lagrangian density of the form
  \be
  \mathcal{L}^{(4)}_{GJP}
 =
 \tfrac{1}{2}\Omega_{\alpha\beta}\wdg*\theta^{\alpha\beta}+f(\mathcal{T})*1,
  \ee
  where $f=f(\mathcal{T})$ is an algebraic function of the topological invariant $\mathcal{T}$.

 In order to find the metric field equations for  $\mathcal{L}_{GJP}$, we introduce as for $\mathcal{L}_{JP}$, the total Lagrangian  $\mathcal{L}^{(4)}_{tot}=  \mathcal{L}_{GJP}+\mathcal{L}_{LM} $. Using (\ref{der2}), the total variational derivative  of the total Lagrangian density can be found as
 \begin{align}
\delta\mathcal{L}^{(4)}_{tot}
&=
\delta\omega_{\alpha\beta}\wdg
\left\{
2D (f'\Omega^{\alpha\beta})
-
\tfrac{1}{2}(\theta^\alpha\wdg \lambda^{\beta}-\theta^\beta\wdg\lambda^{\alpha})
\right\}
+
\delta\lambda_\alpha\wdg \Theta^\alpha
\nonumber\\
&\qquad+
\delta\theta^{\alpha}\wdg
\left\{
\tfrac{1}{2}\Omega^{\mu\nu}\wdg*\theta_{\alpha\mu\nu}
+
(f'\mathcal{T}-f)*\theta_\alpha
+
D\lambda_\alpha
\right\}.
\end{align}
Solving for the Lagrange multiplier 2-forms in terms of the other fields and then inserting them into the equations for coframe 1-forms
we obtain the metric field equations  as
\be\label{GJP}
*G^\alpha
=
(f'\mathcal{T}-f)*\theta^\alpha
+
4Di_\beta (df'\wdg \Omega^{\beta\alpha})
-
\theta^\alpha\wdg Di_\mu i_\nu (df'\wdg \Omega^{\mu\nu}),
\ee
with the condition that $\Theta^\alpha=0$. In the generalization considered here, the metric field equations (\ref{GJP}) remain to be 3$^{rd}$ order in metric components. Note however that,  for the simple choice $f(\mathcal{T})=\mathcal{T}^m$, the first  term on the right hand side is like a cosmological constant proportional to $\mathcal{T}^m$ and the Lagrange multiplier part, or $D\lambda^\alpha$  part, contains $d\mathcal{T}$ instead of the gradient of non-geometrical scalar field $\theta$. In contrast  to the equations of motion of the gravitational model of Jackiw-Pi (\ref{JP}), the first term on the right hand side appears to be an  improvement  term.  Because of this crucial term, the zero trace $(R=0)$ condition for the vacuum metric equations and the original topological constraint $*\mathcal{T}=\Omega^{\alpha\beta}\wdg\Omega_{\alpha\beta}=0$ for the original metric equations for $\mathcal{L}_{JP}$  are to be replaced with
 \be\label{trace-gcs}
 R=4(f-f'\mathcal{T}).
 \ee
This  equation can be found by tracing the first term on the right  hand side in (\ref{GJP}), where the Cotton part does not  contribute to the  trace as in the original Chern-Simons gravitational model of Jackiw-Pi. The new ``constraint equation" (\ref{trace-gcs}) allows  non-zero values of both of the scalars $\mathcal{T}$ and $R$. The scalar $\mathcal{T}$ can be expressed in terms of the components of the Weyl tensor relative to the complex null coframe
of the standard Newman-Penrose formalism. Explicitly in terms of the complex Weyl spinors $\Psi_k$ with $k=0, 1, 2, 3, 4,$ the scalar $\mathcal{T}$ can be written out as
\be
\mathcal{T}
=
4i\left\{3(\bar{\Psi}_2^2-\Psi_2^2)+(\bar{\Psi}_0\bar{\Psi}_4-\Psi_0\Psi_4)+4(\Psi_1\Psi_3-\bar{\Psi}_1\bar{\Psi}_3)\right\},
\ee
where a bar over a quantity denotes complex conjugation. For Petrov type II spaces, it is possible to find a complex null coframe such that $\Psi_0=\Psi_1=0$ with the other Weyl spinors are non-vanishing. Thus, Petrov type II solutions are not allowed in the theory with $\mathcal{T}=0$. For black hole solutions, which are all of Petrov type D, there is a  preferred  complex null coframe for which the only non-vanishing Weyl spinor is $\Psi_2$. In this case, $\mathcal{T}=0$ requires  $\Psi_2$ to be real and therefore, the Petrov D types with real $\Psi_2$, such as Schwarzschild space-time, are allowed whereas the  Kerr solution with complex $\Psi_2$ is ruled out. Finally, Petrov types III, N and  O (conformally flat) one has $\mathcal{T}=0$ identically.

\section{Three Dimensions}

In this last section, we will apply  the metric method presented above to the three dimensional cases, in particular to the massive gravity models involving modified gravitational Lagrangian densities. The presentation in the section also reflects the fact that spacetime dimension $n$ is a crucial parameter in modified gravitational models and demonstrates that  study of the various modified gravitational actions on a case by case basis can be advantageous.

In three spacetime dimensions the Einstein-Hilbert  action (\ref{EH-lag}) does not lead to a dynamical spacetime model \cite{barrow}. Therefore, various higher order models, such as, third order parity-violating  Topologically massive gravity (TMG) \cite{DJT} theory and  New  massive gravity (NMG) theory   proposed recently \cite{bergshoeff}. At the linearized level about Minkowski background,  NMG  can be considered as an extension of Pauli-Fierz  theory to massive spin-2 particle in three spacetime dimensions. In a perturbative approach to quantum gravity, NMG has the remarkable   properties that it is parity invariant, unitary at tree level, ghost-free  \cite{bergshoeff} and renormalizable \cite{oda}. Various classical solutions to NMG theory has been studied. AdS-wave \cite{giribet},  Kundt wave \cite{chakhad} and black hole solutions \cite{bergshoeff2,clement} to  NMG has also been worked out.  A method, based on Killing vectors, is
introduced to study the exact solutions to   three dimensional massive gravity theories \cite{gurses}. In this section we study the field equations of NMG theory and express it in a new form in terms of Cotton 2-form. We will work with  General massive gravity (GMG)  theory which is based on the combination of
TMG and NMG  theories complementing the usual Einstein-Hilbert term \cite{bergshoeff}. Before we investigate  the field equations for General Massive Gravity  Theory, we first recall some geometrical identities for the curvature tensors which are special  to three  spacetime dimensions. In three dimensions, because  the Weyl tensor vanishes identically, the Riemanian curvature can be written in terms of contractions and  in this case, the identity (\ref{weyl-def}) reduces to the form
\be\label{3d-curv-expand}
\Omega^{\alpha\beta}=\theta^{\alpha}\wedge R^{\beta}-\theta^{\beta}\wedge R^{\alpha}
-
\tfrac{1}{2}R\theta^{\alpha\beta}.
\ee
Consequently, the contraction of this identity side by side leads to another  useful identity
\be\label{3d-curv-sqr-id}
\Omega_{\alpha\beta}\wdg *\Omega^{\alpha\beta}
=
2R^{\alpha}\wdg *R_\alpha
-
\tfrac{1}{2}R^2*1,
\ee
which will be exploited below. First and the foremost, as a result of the identity (\ref{3d-curv-sqr-id}), in three spacetime dimensions, the most general Lagrangian density involving  quadratic curvature terms  can   be written either in the form
\be
aR^{\alpha}\wdg *R_\alpha
+
bR^2*1,
\ee
or equivalently in  the alternate form
\be
a\Omega_{\alpha\beta}\wdg *\Omega^{\alpha\beta}
+
(2b+\tfrac{1}{2}a)R^2*1,
\ee
with $a, b$ being arbitrary constants.  As we have noted before, this is also the case for the gravitational  Lagrangians  with quadratic curvature terms in four spacetime dimensions as a result of  a different geometrical identity (cf. Eqn (\ref{des-bay}) in four dimensions). As a simple cross-check for the various  field equations derived in the previous sections, it is possible to obtain the metric field equations for the Lagrangian density
\be\label{3d-equiv-lag1}
\mathcal{L}^{(3)}_{1}
=
\tfrac{1}{2}\Omega_{\alpha\beta}\wdg *\Omega^{\alpha\beta}
+
\tfrac{1}{4}R^2*1,
\ee
from the metric field equations (\ref{ricci-form2}) of the Ricci-squared Lagrangian density
\be\label{3d-equiv-lag2}
\mathcal{L}^{(3)}_{2}
=
R^{\alpha}\wdg *R_\alpha,
\ee
and compare the resulting equations with those that follow directly from (\ref{3d-equiv-lag2}).  In order to do so, inserting the appropriate expressions from Eqns. (\ref{3d-curv-expand}) and (\ref{3d-curv-sqr-id}) into the metric field equations $*\mathcal{P}_\alpha=0$ (see Eqn. (\ref{ricci-form2}) above),
one obtains precisely the metric field equations corresponding to the  Lagrangian density (\ref{3d-equiv-lag1}),  which is  expected as a result of the equivalence of the Lagrangians provided by the identity in (\ref{3d-curv-sqr-id}) in three spacetime dimensions. These considerations are all in accordance with the consistency of  the metric field equations for  both of  the Lagrangians (\ref{3d-equiv-lag1})  and (\ref{3d-equiv-lag2})  which  can  readily
be written down using the general formulae found in the previous sections, namely Eqns. (\ref{f(R)-metric-eqn}), (\ref{ricci-form2}) and (\ref{sky-eqn}).

NMG  also complements  the usual Einstein-Hilbert term with  a particular combination of the Ricci-squared and scalar curvature-squared terms, which
is called $K$-combination \cite{bergshoeff}. GMG is an extension of NMG Lagrangian with a Lorentz Chern-Simons term and a cosmological constant $\Lambda$, and its original form \cite{bergshoeff} can be expressed in terms of differential forms as
\begin{align}
\mathcal{L}^{(3)}_{GMG}
&=
\Lambda*1
-
\tfrac{1}{2}R*1
+
\mathcal{L}^{(3)}_{K}
+
\mathcal{L}^{(3)}_{CS}
\nonumber\\
&=
\Lambda*1
-
\tfrac{1}{2}R*1
+
\tfrac{1}{2m^2}
(R^{\alpha}\wdg *R_\alpha-\tfrac{3}{8}R^2*1)
+
\tfrac{1}{4\mu}\left(\omega^{\alpha}_{\fant{q}\beta}\wdg d\omega^{\beta}_{\fant{q}\alpha}
+
\tfrac{2}{3}\omega^{\alpha}_{\fant{q}\beta}\wdg\omega^{\beta}_{\fant{q}\gamma}\wdg\omega^{\gamma}_{\fant{q}\alpha}\right)\label{gmg-lag1},
\end{align}
where $\mu$ is the mass parameter in TMG theory whereas $m^2$ is the mass parameter in NMG theory. Therefore, the GMG Lagrangian,   which has two mass parameters, reduces to that of TMG Lagrangian in the limit $m\mapsto\infty$ and to that of NMG Lagrangian  in the limit $\mu\mapsto\infty$. Note that the GMG Lagrangian density (\ref{gmg-lag1}) can equivalently be written in the form
\be\label{gmg-lag2}
\mathcal{L}^{(3)}_{GMG}
=
\Lambda*1
-
\tfrac{1}{2}R*1
+
\tfrac{1}{4m^2}
(
\Omega_{\alpha\beta}\wdg *\Omega^{\alpha\beta}
-
\tfrac{1}{4}R^2*1)
+
\tfrac{1}{4\mu}\left(\omega^{\alpha}_{\fant{q}\beta}\wdg d\omega^{\beta}_{\fant{q}\alpha}
+
\tfrac{2}{3}\omega^{\alpha}_{\fant{q}\beta}\wdg\omega^{\beta}_{\fant{q}\gamma}\wdg\omega^{\gamma}_{\fant{q}\alpha}\right),
\ee
by using the identity given in (\ref{3d-curv-sqr-id}). The variational derivative of the Lagrangian density (\ref{gmg-lag1}) (or equivalently, (\ref{gmg-lag2}))
 and the resulting Lagrange multiplier 1-forms have special properties that deserve further scrutiny.  We choose to work with the form (\ref{gmg-lag2}) and
the constrained total variation derivative of the Lagrangian density $\mathcal{L}_{tot}=\mathcal{L}^{(3)}_{GMG}+\mathcal{L}_{LM}$ leads to
\begin{align}
\delta\mathcal{L}_{tot}
&=
\delta \omega_{\alpha\beta}\wdg
\left\{
\tfrac{1}{2m^2}D*(\Omega^{\alpha\beta}-\tfrac{1}{4}R\theta^{\alpha\beta})
+
\tfrac{1}{2\mu}\Omega^{\alpha\beta}
-
\tfrac{1}{2}D*\theta^{\alpha\beta}
-
\tfrac{1}{2}(\theta^{\alpha}\wdg \lambda^\beta-\theta^{\beta}\wdg \lambda^\alpha)
\right\}
\nonumber\\
&+
\delta \theta_{\alpha}\wdg
\left\{
(1+\tfrac{R}{4m^2})*G^\alpha
+
(\Lambda+\tfrac{R^2}{16m^2})*\theta^\alpha
+
\tfrac{1}{2m^2}*T^\alpha[\Omega]+D\lambda^\alpha
\right\}
+
\delta\lambda_\alpha\wdg \Theta^\alpha,
\nonumber
\end{align}
(up to an omitted closed form) where the definition of the canonical energy momentum 2-form for curvature 2-forms defined in Eqn. (\ref{en-mom-curv-2form}) has been used for convenience as before.  Note here that the trace of the energy momentum 2-form (\ref{en-mom-curv-2form}) in three dimensions  is
\be
T^{\alpha}_{\fant{a}\alpha}[\Omega]*1=-\tfrac{1}{2}\Omega_{\mu\nu}\wdg*\Omega^{\mu\nu}.
\ee
The Lagrange multiplier 1-form can be solved in terms of other fields using the field equations obtained from the connection variation using general formulae (\ref{general-lag-mult}) and (\ref{3d-curv-expand}). This yields
\be\label{gmg-lag-mul}
\lambda^\beta
=
\tfrac{1}{m^2}
i_\alpha D*(\Omega^{\alpha\beta}-\tfrac{1}{4}R\theta^{\alpha\beta})
+
\tfrac{1}{\mu}(R^\beta-\tfrac{1}{4}R\theta^\beta).
\ee
Here, the first term is the contribution of  $\mathcal{L}_K$ terms and are fourth order in metric components, whereas the last term contributes  to the metric field equations as the Cotton 2-form $C^\alpha=D(R^\alpha-\tfrac{1}{4}R\theta^\alpha)$ and is third order in metric components. Although Cotton 2-form is defined in arbitrary dimensions and $C_\alpha\ot \theta^\alpha$ is invariant under conformal transformation, its vanishing is necessary and sufficient condition for conformal flatness in three dimensions. In addition,   the Cotton 2-form is traceless, $i_\alpha C^\alpha=0$, satisfies Bianchi type identity, $\theta_\alpha\wdg C^\alpha=0$ and  $C^\alpha$ is  covariantly constant,  $DC^\alpha=0$, in three dimensions \cite{cotton}.

The significance of the  particular choice of the Lagrangian density $\mathcal{L}_K$   in (\ref{gmg-lag2}) shows up in the expression for the Lagrange multiplier  terms, namely $D\lambda^\beta$. The first of the terms in (\ref{gmg-lag-mul}) is NMG part whereas the second term  is  the TMG part. The traces of each parts vanishes separately, and therefore they do not contribute to the trace of the metric field equations of the GMG theory, to put it mathematically
one has $\theta_\beta \wdg D\lambda^\beta=0$ for the Lagrange multiplier given  in (\ref{gmg-lag-mul}). Consequently, as can be explicitly verified from the resulting field equations, the trace of the terms in the field equation coming  from the Lagrangian density $\mathcal{L}_K$ add up to  $\mathcal{L}_K$ up to a sign. (We refer to the trace formulae given in Eqns. (\ref{trace-g}), (\ref{ricci-trace}) and  (\ref{sky-trace})).  Using the result (\ref{gmg-lag-mul}), the GMG metric field  equations can be written as
\begin{align}
&(1+\tfrac{1}{4m^2}R)*G^\alpha
+
(\Lambda+\tfrac{1}{16m^2}R^2)*\theta^\alpha
+
\tfrac{1}{\mu}C^\alpha
+
\tfrac{1}{2m^2}*T^\alpha[\Omega]
\nonumber\\
&
-
\tfrac{1}{4m^2}D*(dR\wdg\theta^\alpha)
+
\tfrac{1}{m^2}
Di_\beta D*\Omega^{\beta\alpha}=0.\label{gmg-reduced}
\end{align}
As we shall exemplify below, the metric field equations given in (\ref{gmg-reduced}) for the GMG field equations seem to be more advantageous compared to the original form of the field equations, since for example, for the fact that it can be also expressible in parts or as a whole in terms of Ricci  1-forms  and scalar curvatures using (\ref{3d-curv-expand}). Note that the TMG theory is a third order theory whereas  NMG is a fourth order theory. Using (\ref{3d-curv-expand}) and the fact that in three dimensions  the Ricci 1-forms and scalar curvature satisfy the identity  $i_\alpha DR^\alpha+\tfrac{1}{2}dR=0$  (This identity can be obtained  using the  property that Cotton 2-form is traceless) one finds that the contribution of the Lagrange multiplier term, i.e.  $D\lambda^\beta $ term, to the metric field equations  takes  the form
\be\label{sqrt}
D\lambda^\alpha
=
\tfrac{1}{m^2}
D*C^\alpha
+
\tfrac{1}{\mu}C^\alpha.
\ee
This form of the Lagrangian multiplier also makes it easier to see the various properties of the NMG theory in comparison to TMG. The first term on the right hand side of (\ref{sqrt}), which is fourth order in metric components, is unique with the properties: (i) Trace-free, (ii) Covariantly constant, (iii) Parity-preserving. In contrast,   the  TMG part of the Lagrange multiplier shares these properties except that it is parity-violating.

It is shown that TMG is ``square root" of the NMG theory at the linearized level \cite{bergshoeff}. The presentation of the field equations above makes it  possible to sharpen this result. In fact, such a relation between the two theory  is encoded in the Lagrange multiplier structure of the GMG theory
and  can be put in  mathematically precise  form under certain assumptions using Eqns. (\ref{gmg-reduced}) and (\ref{sqrt}). If we assume for simplicity  that $\Lambda=0$  and $R=0$,  then using (\ref{gmg-reduced}) the TMG and the NMG field equations  can be written as
\begin{align}
&(*D+\mu)R^\alpha=0,\label{tmg2}\\
&(*D*D+m^2)R^\alpha=-\tfrac{1}{2}T^{\alpha}[\Omega],\label{nmg2}
\end{align}
respectively. Eqn. (\ref{tmg2}) can be derived  by simply applying Hodge dual operator $*$  to the TMG equations obtained from  (\ref{gmg-reduced}).
The energy momentum 1-form $T^\alpha[\Omega]$ in the right hand side of Eqn. (\ref{nmg2}) can be expressed in tems of Ricci-squared terms for $R=0$.
This correspondence amounts to the fact that both TMG and NMG can be written in Dirac type and Klein-Gordon type equations respectively. This point of view is  elaborated   in \cite{aliev1} relative to a coordinate coframe  where the correspondence is utilized to provide a method to generate  exact solutions  for the  NMG theory for  any given Petrov-Segre type D and  N solutions of the TMG theory \cite{barrow,cotton,sezgin, TMG-cotton-note}. In particular, the non-dispersive, line-fronted gravitational wave metric  of the Kerr-Schild form
\be\label{pp-wave}
g=2H(u,x)du\ot du+dv\ot du +dv\ot du+dx\ot dx
\ee
expressed in terms of local coordinates  $\{u,v,x\}$ where $u,v$ are real null coordinates and $x$ is a spatial coordinate. The metric (\ref{pp-wave})
linearizes the TMG equation (\ref{tmg2}) for the real metric function $H(u,x)$ \cite{dereli-tucker-3d}. Moreover, for the Petrov-Segre type N metric  (\ref{pp-wave}), one has  $T^\alpha[\Omega]=0$ identically and therefore, it also linearizes the NMG equations (\ref{nmg2}). Consequently,
there is a class of gravitational wave solutions of the form (\ref{pp-wave}) which is common to both TMG and NMG theories.

\section{Concluding Remarks}

We have presented  a unified and practical method to  obtain field equations for some modified gravitational Lagrangians containing quadratic curvature invariants and their algebraic generalizations in $n\geqslant 3$  spacetime dimensions. We have adopted the language of differential forms throughout our presentation which provides  further insights even for  the well-known results. As clearly indicated by our presentation for the GMG theory  in three dimensions relative to an
orthonormal coframe, the derivation of the metric field equations and the Lagrange multiplier terms provides alternative   forms for the field equations and insight  for the structure of them.

Although we did not address the question of coupling matter to specific gravitational model, our treatment can be  even extended to the case to in which
spinor matter fields present or to the case where non-metricity is present. We presented the variational derivatives   with respect to independent coframe and connection 1-forms in an explicit form  for all the generalized  gravitational actions  such that these extensions can easily be pursued. The presentation  also makes it possible to compare the Palatini formalism with the metric formalism. Rather then providing an exhaustive list  of modified gravitational Lagrangians and their corresponding field equations we presented a unified method of the derivation for the field equations using popular gravitational models.
We leave the further studies such as the exact solutions, and other physical and mathematical features  of the new models we put forward, i.e., modified  Chern-Simons theory and GMG theory to forthcoming research.

\section*{Appendix}
In this section we will briefly explain the  notation. For further  details we refer to, for example, \cite{thirring} and \cite{benn-tucker}. The definitions of basic geometrical quantities we use can be summarized  as follows. The metric relative to an orthonormal coframe 1-forms $\theta^\alpha$ is  $g=\eta_{\alpha\beta}\theta^\alpha\ot\theta^\beta$, the wedge products of basis 1-forms are abbreviated as   $\theta^{\alpha\beta\cdots}=\theta^{\alpha}\wdg\theta^{\beta}\wdg\cdots$. The contraction with respect to  frame fields $e_\alpha$ is denoted by $i_\alpha$ where $i_\alpha\theta^\beta=\delta^{\alpha}_{\beta}$. Oriented volume element can be written as $*1=\theta^{01\cdots (n-1)}$ where  $*$ is the linear Hodge dual operator  and $**=(-1)^{s+p(n-p)}$ acting on a $p$-form in an $n$ dimensional spacetime ($s$ is the metric signature). The  indices of the permutation symbol  $\varepsilon_{\alpha_1\alpha_2\cdots\alpha_{n}}$ relative to an orthonormal basis is raised and lowered by $\eta_{\alpha\beta}=\eta^{\alpha\beta}$ and $\varepsilon_{0123\cdots (n-1)}=+1$. Cartan's first structure equations can be used to define  torsion 2-form  and  torsion tensor as
\[
\Theta^{\alpha}=\tfrac{1}{2}T^{\alpha}_{\fant{a}\mu\nu}\theta^{\mu\nu}
=
D\theta^\alpha
=
d\theta^\alpha+\omega^{\alpha}_{\fant{a}\beta}\wdg\theta^\beta,
\]
where $\omega^{\alpha}_{\fant{a}\beta}=\omega^{\alpha}_{\lambda\beta}\theta^\lambda$ are connection 1-forms. The contorsion 1-form is defined as
$\Theta^\alpha=K^{\alpha}_{\fant{a}\beta}\wdg \theta^\beta$. The curvature 2-forms in terms of Riemann tensor  are
\[
\Omega^{\alpha}_{\fant{a}\beta}=\tfrac{1}{2}R^{\alpha}_{\fant{a}\beta\mu\nu}\theta^{\mu\nu}
=
d\omega^{\alpha}_{\fant{a}\beta}+\omega^{\alpha}_{\fant{a}\lambda}\wdg \omega^{\lambda}_{\fant{a}\beta}
\]
(Cartan's second structure equations). Ricci 1-forms can be  defined as the contraction of the curvature 2-forms
\[
R_\beta
\equiv
i_\alpha\Omega^{\alpha}_{\fant{a}\beta}
=
\tfrac{1}{2}R^{\alpha}_{\fant{a}\beta\mu\nu}i_\alpha(\theta^{\mu}\wdg\theta^{\nu})
=
R^{\mu}_{\fant{a}\beta\mu\lambda}\theta^\lambda
=
R_{\beta\lambda}\theta^\lambda\]
where
$R_{\beta\lambda}$ are the components of the Ricci tensor. The scalar curvarture is $R=i_\alpha R^\alpha$. Similarly,
the components of the Weyl 2-form $C^{\alpha}_{\fant{a}\beta}$ and  those of the Weyl tensor $C^{\alpha}_{\fant{a}\beta\mu\nu}$ are related by
$C^{\alpha}_{\fant{a}\beta}=\tfrac{1}{2}C^{\alpha}_{\fant{a}\beta\mu\nu}\theta^{\mu\nu}$.

 A tensor $\phi$ of type $(r+p,s)$ which is totally antisymmetric  with respect to $p$ number of indices can uniquely be associated with the $(r,s)$ tensor-valued $p$-form
$\phi^{\alpha_1\alpha_2\cdots \alpha_s}_{\fant{aaaaaaaa}\beta_1\beta_2\cdots \beta_r}$. Relative to an arbitrary coframe
considered as multilinear maps on (co)frame fields they are related  by
\[
\phi^{\alpha_1\alpha_2\cdots \alpha_s}_{\fant{aaaaaaaa}\beta_1\beta_2\cdots \beta_r}(e_{\beta_1},e_{\beta_2},\cdots, e_{\beta_p})
\equiv
\phi(\theta^{\alpha_1},\theta^{\alpha_2},\cdots,\theta^{\alpha_s}, e_{\beta_1},e_{\beta_2},\cdots, e_{\beta_r}, e_{\beta_1},e_{\beta_2},\cdots e_{\beta_p} ).
\]
Covariant exterior derivative $D$  acts on tensor-valued $p$-forms to yield tensor-valued $(p+1)$-form preserving the tensorial type as follows
\[
D\phi^{\alpha_1\alpha_2\cdots \alpha_s}_{\fant{aaaaaaaa}\beta_1\beta_2\cdots \beta_r}
=
d\phi^{\alpha_1\alpha_2\cdots \alpha_s}_{\fant{aaaaaaaa}\beta_1\beta_2\cdots \beta_r}
+
\omega^{\alpha_1}_{\fant{a}\lambda}\wdg\phi^{\lambda\alpha_2\cdots \alpha_s}_{\fant{aaaaaaaa}\beta_1\beta_2\cdots \beta_r}
+
\cdots
-
\omega^{\lambda}_{\fant{a}\beta_1}\wdg \phi^{\alpha_1\alpha_2\cdots \alpha_s}_{\fant{aaaaaaaa}\lambda \beta_2\cdots \beta_r}
-\cdots
\]
More generally, for  tensor valued  $p$-form $\phi^I$ and tensor valued $q$-form $\psi^{J}$
\[
D(\phi^I\wdg \psi^J)
=
D\phi^I\wdg \psi^J+(-1)^p\phi^I\wdg D\psi^J.
\]
Acting on scalars $D$ reduces to exterior derivative $d$.
The expression
$D\phi^{\alpha_1\alpha_2\cdots \alpha_s}_{\fant{aaaaaaaa}\beta_1\beta_2\cdots \beta_r}$ can easily be related the covariant derivative $\nabla_X\phi$ using the relation between the exterior and Riemannian covariant derivatives, namely $d=\theta^\alpha\wdg \nabla_\alpha$. The covariant derivative
$\nabla_X=X^\alpha\nabla_\alpha$ can be defined by its action on basis coframe:  $\nabla_X\theta^{\alpha}=-(i_X\omega^{\alpha}_{\fant{a}\beta})\theta^\beta$.


\begin{thebibliography}{99}
\bibitem{TMG-cotton-note} The Cotton tensor is classified (into Petrov types) according to the eigenvalues of traceless
$3\times 3$ matrix $C_{\alpha\beta}=C_{\beta\alpha}$ which can be defined as $(C_{\alpha\beta})=i_\beta*C_\alpha$ in terms of Cotton 2-form.
The TMG field equations render the classification of traceless Ricci  tensor (into Segre types) and  Petrov type equivalent.
\end{thebibliography}
\end{document}